\documentclass[default,iicol]{sn-jnl}

\jyear{2021}%

\theoremstyle{thmstyleone}%
\theoremstyle{thmstyletwo}%

\theoremstyle{thmstylethree}%

\usepackage{siunitx}
\usepackage{booktabs}
\usepackage{subcaption}
\usepackage{amsmath}
\usepackage{microtype}
\usepackage{lineno}  
\usepackage{xspace} 
\usepackage{caption} 

\usepackage{graphicx}  
\usepackage{color}
\usepackage{colortbl}
\graphicspath{{./figs/}} 

\usepackage{amsmath} 
\usepackage{amssymb}
\usepackage{amsfonts}
\usepackage{upgreek} 

\usepackage{hyperxmp}
\usepackage[all]{hypcap} 
\usepackage{listings}
\usepackage{ifthen} 
\newboolean{uprightparticles}
\setboolean{uprightparticles}{false} 

\usepackage{float} 

\usepackage[numbers]{natbib} 

\usepackage{xspace} 
\usepackage{upgreek}

\def\lhcb   {\mbox{LHCb}\xspace}

\def\MagUp {\mbox{\em Mag\kern -0.05em Up}\xspace}

\def\hltone {HLT1\xspace}
\def\hlttwo {HLT2\xspace}

\ifthenelse{\boolean{uprightparticles}}%
{

 \def\PDelta      {\ensuremath{\Delta}\xspace}                 
 \def\PXi      {\ensuremath{\Xi}\xspace}                 
 \def\PLambda      {\ensuremath{\Lambda}\xspace}                 
 \def\PSigma      {\ensuremath{\Sigma}\xspace}                 
 \def\POmega      {\ensuremath{\Omega}\xspace}                 
 \def\PUpsilon      {\ensuremath{\Upsilon}\xspace}                 
 
 \def\PB      {\ensuremath{\mathrm{B}}\xspace}                 
                  
 \def\PD      {\ensuremath{\mathrm{D}}\xspace}

 \def\PK      {\ensuremath{\mathrm{K}}\xspace}

 \def\Pi      {\ensuremath{\mathrm{i}}\xspace}

}
{

 \mathchardef\PDelta="7101
 \mathchardef\PXi="7104
 \mathchardef\PLambda="7103
 \mathchardef\PSigma="7106
 \mathchardef\POmega="710A
 \mathchardef\PUpsilon="7107
                  
 \def\PB      {\ensuremath{B}\xspace}                 
                  
 \def\PD      {\ensuremath{D}\xspace}

 \def\PK      {\ensuremath{K}\xspace}

 \def\Pi      {\ensuremath{i}\xspace}

}

\makeatletter
\ifcase \@ptsize \relax
  \newcommand{\miniscule}{\@setfontsize\miniscule{4}{5}}
\or
  \newcommand{\miniscule}{\@setfontsize\miniscule{5}{6}}
\or
  \newcommand{\miniscule}{\@setfontsize\miniscule{5}{6}}
\fi
\makeatother

\DeclareRobustCommand{\optbar}[1]{\shortstack{{\miniscule (\rule[.5ex]{1.25em}{.18mm})}
  \\ [-.7ex] $#1$}}

\def\kaon    {{\ensuremath{\PK}}\xspace}
  \def\Kbar    {{\kern 0.2em\overline{\kern -0.2em \PK}{}}\xspace}

\def\KorKbar    {\kern 0.18em\optbar{\kern -0.18em K}{}\xspace}

\def\KS      {{\ensuremath{\kaon^0_{\mathrm{ \scriptscriptstyle S}}}}\xspace}

  \def\Dbar    {{\kern 0.2em\overline{\kern -0.2em \PD}{}}\xspace}

\def\DorDbar    {\kern 0.18em\optbar{\kern -0.18em D}{}\xspace}

\def\Bbar    {{\ensuremath{\kern 0.18em\overline{\kern -0.18em \PB}{}}}\xspace}

\def\BorBbar    {\kern 0.18em\optbar{\kern -0.18em B}{}\xspace}

  \def\Y#1S{\ensuremath{\PUpsilon{(#1S)}}\xspace}

\def\Lbar        {{\ensuremath{\kern 0.1em\overline{\kern -0.1em\PLambda}}}\xspace}
\def\LorLbar    {\kern 0.18em\optbar{\kern -0.18em \PLambda}{}\xspace}

\def\to                 {\ensuremath{\rightarrow}\xspace}

\def\AT#1     {\ensuremath{A_{\mathrm{T}}^{#1}}\xspace}           

\def\C#1      {\ensuremath{\mathcal{C}_{#1}}\xspace}                       
\def\Cp#1     {\ensuremath{\mathcal{C}_{#1}^{'}}\xspace}                    
\def\Ceff#1   {\ensuremath{\mathcal{C}_{#1}^{\mathrm{(eff)}}}\xspace}        
\def\Cpeff#1  {\ensuremath{\mathcal{C}_{#1}^{'\mathrm{(eff)}}}\xspace}       
\def\Ope#1    {\ensuremath{\mathcal{O}_{#1}}\xspace}                       
\def\Opep#1   {\ensuremath{\mathcal{O}_{#1}^{'}}\xspace}                    

\newcommand{\tev}{\ifthenelse{\boolean{inbibliography}}{\ensuremath{~T\kern -0.05em eV}}{\ensuremath{\mathrm{\,Te\kern -0.1em V}}}\xspace}
\newcommand{\gev}{\ensuremath{\mathrm{\,Ge\kern -0.1em V}}\xspace}
\newcommand{\mev}{\ensuremath{\mathrm{\,Me\kern -0.1em V}}\xspace}
\newcommand{\kev}{\ensuremath{\mathrm{\,ke\kern -0.1em V}}\xspace}
\newcommand{\ev}{\ensuremath{\mathrm{\,e\kern -0.1em V}}\xspace}
\newcommand{\gevc}{\ensuremath{{\mathrm{\,Ge\kern -0.1em V\!/}c}}\xspace}
\newcommand{\mevc}{\ensuremath{{\mathrm{\,Me\kern -0.1em V\!/}c}}\xspace}
\newcommand{\gevcc}{\ensuremath{{\mathrm{\,Ge\kern -0.1em V\!/}c^2}}\xspace}
\newcommand{\gevgevcccc}{\ensuremath{{\mathrm{\,Ge\kern -0.1em V^2\!/}c^4}}\xspace}
\newcommand{\mevcc}{\ensuremath{{\mathrm{\,Me\kern -0.1em V\!/}c^2}}\xspace}

\def\mm   {\ensuremath{\mathrm{ \,mm}}\xspace}

\def\gsim{{~\raise.15em\hbox{$>$}\kern-.85em
          \lower.35em\hbox{$\sim$}~}\xspace}
\def\lsim{{~\raise.15em\hbox{$<$}\kern-.85em
          \lower.35em\hbox{$\sim$}~}\xspace}

\def\pt         {\ensuremath{p_{\mathrm{ T}}}\xspace}

\def\tell1  {TELL1\xspace}
\def\ukl1   {UKL1\xspace}

\usepackage{subfiles} 

\newcommand{\HS}{\texttt{Seeding}\xspace}
\newcommand{\Forward}{\texttt{Forward}\xspace}
\newcommand{\matching}{\texttt{Matching}\xspace}
\newcommand{\seeding}{\texttt{Seeding}\xspace}
\newcommand{\VELO}{VELO\xspace}
\newcommand{\scifi}{SciFi\xspace}
\newcommand{\standalone}{standalone\xspace}

\newcommand{\longtrack}{Long\xspace}

\newcommand{\xz}{$x$-$z$\xspace}
\newcommand{\zref}{\ensuremath{z_{\rm ref}}\xspace}

\newcommand{\runIII}{Run~3\xspace}
\newcommand{\zmatchx}{\ensuremath{z_x^{\rm match}}\xspace}
\newcommand{\zmatchy}{\ensuremath{z_y^{\rm match}}\xspace}
\newcommand{\tolx}{\ensuremath{\mathrm{tol}_x}\xspace}
\newcommand{\toly}{\ensuremath{\mathrm{tol}_y}\xspace}
\newcommand{\toltx}{\ensuremath{\mathrm{tol}_{t_x}}\xspace}
\newcommand{\tolty}{\ensuremath{\mathrm{tol}_{t_y}}\xspace}
\newcommand{\chisqmatch}{\ensuremath{\chi^2_{\rm match}}\xspace}
\newcommand{\zb}[1]{\ensuremath{\bar{z}_{#1}}\xspace}
\newcommand{\dratio}{\ensuremath{d_{\rm ratio}}\xspace}

\begin{document}


\title[Article Title]{\HS and
\matching algorithms for the first
GPU-based High Level Trigger of the
LHCb experiment}


\author*[1]{\fnm{C.} \sur{Agapopoulou}}\email{christina.agapopoulou@cern.ch}
\author*[2]{\fnm{L.} \sur{Calefice}}\email{lukas.calefice@cern.ch}
\author[3]{\fnm{A.} \sur{Fernández~Casani}}
\author[1]{\fnm{V.} \sur{Gligorov}}
\author[4]{\fnm{A.} \sur{Hennequin}}
\author[5]{\fnm{L.} \sur{Henry}}
\author*[3,5]{\fnm{V.} \sur{Kholoimov}}\email{valerii.kholoimov@cern.ch}
\author[6]{\fnm{B.} \sur{Kishor Jashal}}
\author[3]{\fnm{A.} \sur{Oyanguren Campos}}
\author*[7]{\fnm{L.} \sur{Pica}}\email{lorenzo.pica@cern.ch}
\author[3]{\fnm{V.} \sur{Svintozelskyi}}
\author[8]{\fnm{D.Y.} \sur{Tou}}
\author*[3]{\fnm{J.}\sur{Zhuo}}\email{jiahui.zhuo@cern.ch}

\affil[1]{LPNHE, Sorbonne Université, CNRS/IN2P3, Paris, France }
\affil[2]{Universitat de Barcelona, Barcelona, Spain} 
\affil[3]{Instituto de Física Corpuscular (IFIC), University of Valencia-CSIC, Valencia, Spain}
\affil[4]{CERN, European Organization for Nuclear Research, Switzerland}
\affil[5]{Institute of Physics,
\'Ecole Polytechnique F\'ed\'erale de Lausanne (EPFL), Lausanne, Switzerland}
\affil[6]{Rutherford Appleton Laboratory (RAL), Oxford, United Kingdom}
\affil[7]{Scuola Normale Superiore $\&$ INFN Pisa, Italy}
\affil[8]{Tsinghua University, Beijing, China}


\abstract{We describe the GPU implementation of the \HS and \matching algorithms, developed for the first level trigger of the LHCb experiment and key to reconstruct long and very displaced tracks at 40~MHz. The algorithms have been participating in the data taking during the full Run~3 of LHCb with a very high throughput, increasing the physics reach of the experiment. 

The \HS is a \standalone pattern recognition algorithm aiming at finding charged particle trajectories in the most forward tracker of LHCb.
These trajectories are then extrapolated backward by the \matching algorithm which combines them with stubs formed from hits in the first tracker 
in order to form what we call \longtrack tracks.
Hits in the second tracker
are then searched for to better define the trajectory and improve the track momentum resolution. 
This backward approach, complementary to the approach of extrapolating the 
stubs in the first detector to the forward tracker through the magnetic field, improves the \longtrack track efficiency at low transverse momenta, increasing the potential of key physics decay channels.}

\keywords{LHCb, trigger, HLT1, seeding, GPU, LLPs}



\maketitle

\section{Introduction}\label{sec:introduction}

The LHCb detector at the LHC~\cite{Aaij_2024} has undergone a major upgrade in preparation of the Run~3 data-taking, which started in 2022. 
The detector operates at present an instantaneous luminosity of $\mathcal{L} = \SI{2e33}{cm^{-2}s^{-1}}$, corresponding to an average of five proton-proton interactions per bunch collision. 
The entire charged particle reconstruction (tracking) system of the \lhcb detector has been renewed as part of this upgrade. In particular, the system closest to the interaction point, the VErtex LOcator (VELO), has been upgraded from a silicon strip to a silicon pixel design, allowing for 3-dimensional position information 
described in detail in Ref.~\cite{Campora:search_by_triplet}. 
The Upstream Tracker (UT), located in front of the entrance of the LHCb dipole magnet, was also redesigned  and is now composed
of four planes of higher-granularity silicon-strip detectors. Finally, the tracker placed downstream of the \lhcb dipole magnet was replaced by a scintillating fibre tracker (\scifi) described in detail in Ref.~\cite{LHCb:2014uqj}. 
The \scifi consists of three stations (T1, T2, T3), each composed of four layers of stacked scintillating fibres. The layers within one station are separated from each other by an air-filled gap of \SI{50}{\milli\meter}. 
The LHCb dipole
magnet has a bending power of 4 Tm and the magnetic field vector is oriented vertically, hence tracks are bent mostly in the horizontal plane ($x-z$).
The polarity of the LHCb dipole magnet is periodically reversed between magnet-up and magnet-down configurations to reduce detector-induced charge asymmetries.
Both the UT and \scifi  layers are arranged in a stereo configuration ($x$-$u$-$v$-$x$), to have measurements in the bending ($x$) as well as in the non-bending ($y$) direction.
The $x$-layers are vertically oriented while the $u/v$ layers are rotated in the $x-y$ plane by the stereo angle, $\alpha$, equal to $+5^{\circ}$ and $-5^{\circ}$ for the $u$ and $v$ layers, respectively. 

The \lhcb trigger system has also undergone a major upgrade between Run 2 and Run 3, with the hardware stage being completely removed and the detector being read-out at the full bunch crossing rate of 40\,MHz. The event selection strategy relies on two High Level Trigger (HLT) software stages called \hltone and \hlttwo~\cite{CERN-LHCC-2014-016}.
The \hltone stage performs a partial reconstruction of the event focusing on selecting displaced and significant transverse momentum signatures. 
As there is no hardware trigger available, track reconstruction becomes the crucial aspect of the trigger with a rate of 2 billion tracks per second, a factor 20 larger as compared to other LHC experiments such as ATLAS or CMS.  
The HLT1 trigger is fully implemented on commercial NVIDIA A5000 GPUs and reduces the input data rate of 40\,Tbit/s by a factor of 30. 
More details can be found in Ref.~\cite{Allen}, where the project handling all HLT1 algorithms, \texttt{Allen}, is described.
The \hlttwo reconstruction exploits the reduced event rate after the \hltone selections to perform a full reconstruction of the event, adding offline-quality track reconstruction and particle identification information. This reconstruction also benefits from the real-time alignment and calibration procedure developed for the Run~2 data taking~\cite{trigger_performance_run2} and adapted for Run~3.

\begin{figure}[h]
    \centering
    \includegraphics[width=0.5\textwidth]{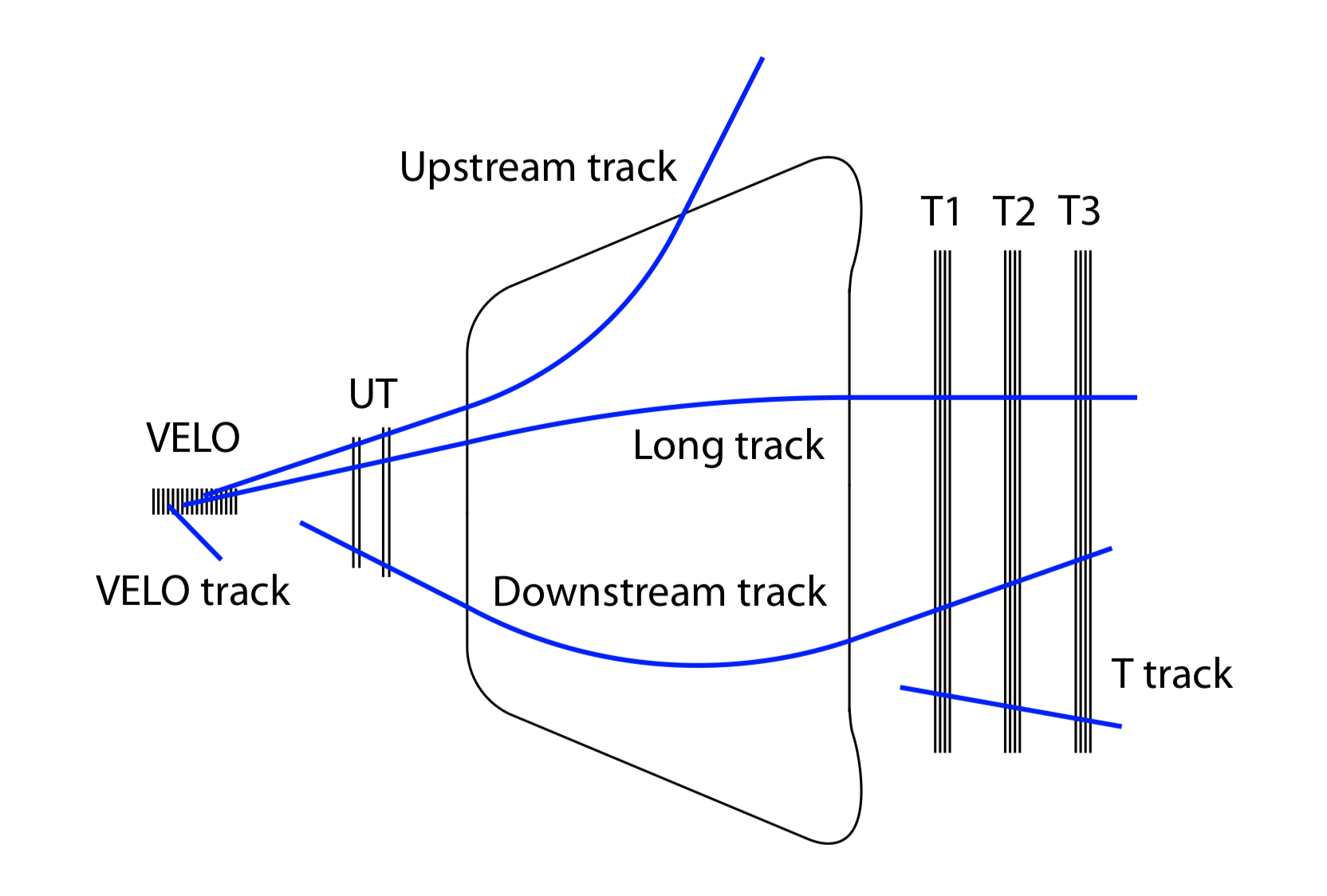}
    \caption{Schematic of track types in the LHCb detector.}
    \label{fig:track_types}
\end{figure}

The different types of tracks reconstructed in LHCb are shown in Fig.~\ref{fig:track_types}. The majority of trigger decisions are based on Long tracks, originating from particles that leave hits in all three tracking sub-detectors.  In \hlttwo, Long tracks are reconstructed based on two approaches, the \Forward and the \HS algorithms. The former is based on extending VELO track segments to the \scifi stations and searching for compatible hits. The general design was kept but the algorithm has been tuned for Run~3~\cite{Gunther:2023wwm}. On the other hand, the \HS approach~\cite{HybridSeeding} reconstructs \scifi segments, or \textit{T} tracks, through the \HS algorithm, and then matches them to the \VELO tracks through the \matching algorithm~\cite{Needham:2007oba,Needham:2007zzb}. The two approaches offer important complementarity and redundancy while being within the timing requirements of \hlttwo, they therefore run independently of each other. The stronger throughput limitations of \hltone led to the adoption of only the \Forward algorithm as the default track reconstruction in Run~2~\cite{trigger_performance_run2}.

This paper describes the first GPU adaptation of the backward reconstruction of Long tracks based on the \HS and \matching algorithms, and their successful implementation to reconstruct about two billion tracks per second in the \hltone trigger of LHCb in Run~3.
The implementation of the GPU-based \HS algorithm has enabled the implementation of the \texttt{Downstream} algorithm~\cite{Kholoimov:2025cqe}, largely increasing the physics potential of the LHCb experiment for detecting particles of long lifetime \cite{Gorkavenko_2024}.

The paper is organized as follows: Section~\ref{sec:motivation} describes the motivation for this work, while Sections~\ref{sec:hseeding} and ~\ref{sec:matching} give details on the \HS and \VELO-\scifi \matching algorithms. The key challenges for the GPU implementation is explained in Section~\ref{sec:gpus}. 
The performance of the algorithms within the context of the \hltone physics and timing requirements is then presented in Section~\ref{sec:performance}, followed by the physics impact in  Section~\ref{sec:physics}.
Conclusions and future prospects are given in Section~\ref{sec:conclusion}.

\section{Motivation}\label{sec:motivation}

Two main reasons motivate the algorithm development of this work: enabling the downstream reconstruction at \hltone and improving the overall efficiency for Long tracks.  
As discussed in Section ~\ref{sec:introduction}, the majority of LHCb's trigger selections are based on Long tracks. Downstream tracks, which only contain hits from the UT and \scifi detectors, offer a lower \pt and impact parameter resolution with respect to Long tracks, but are crucial to extend detector acceptance to those long-lived particles decaying outside the \VELO region. Such particles can be \KS mesons and $\Lambda$ baryons, but also potential New Physics (NP) candidates. The reconstruction of Downstream tracks is based on extending T tracks from the \HS algorithm towards the UT. During Run~2 this reconstruction was only available at the \hlttwo stage of LHCb. The lack of an equivalent algorithm for HLT1 significantly reduced the experiment sensitivity to any signal mode involving \KS or $\Lambda$ hadrons, as well as long-lived NP signatures, as was outlined in Ref.~\cite{SeedingLLP}.
Thanks to the implementation of the \HS algorithm in the \hltone level that is described in this paper, the reconstruction of Downstream tracks at HLT1 became accessible for the first time in LHCb. It has extended the physics reach of the LHCb in Run~3 for several interesting channels involving long-lived particles in their final state. At the same time, it has allowed to test and to acquire experience on Downstream track-based trigger selections in view of Run~4, when LHCb is planning to install a FPGA-based device (the so-called Downstream Tracker)~\cite{LHCb-tdr-025, Morello:2888549}. It aims at bringing the \HS algorithm timing close to zero by solving the pattern recognition task before the event building step, freeing resources at the HLT1 level that can be spent to improve reconstruction and selections performances.

In the baseline \hltone \Forward tracking algorithm, only Long tracks with transverse momentum above \SI{450}{MeV/c} are reconstructed. This limitation is imposed in order to satisfy the \hltone throughput requirements, since the search for \scifi hits for low momentum tracks would need to be performed over a much larger area of the detector than for higher-momentum tracks. However, this limitation leads to a reduced sensitivity of HLT1 to low-momentum signatures such as the ones originating from the decays of charm and strange particles. In the \HS and \matching reconstruction approach described in this paper, this limitation is effectively lifted since the new algorithms do not need to impose stringent momentum or transverse momentum criteria.

\def\alphacorr{\ensuremath{\alpha_{\rm corr}}\xspace}
\def\qop{\ensuremath{q/p}\xspace}
\def\ax{\ensuremath{a_x}\xspace}
\def\bx{\ensuremath{b_x}\xspace}
\def\cx{\ensuremath{c_x}\xspace}
\section{The \HS algorithm in GPUs}
\label{sec:hseeding}

The \HS algorithm was originally developed for the \hlttwo, named \hlttwo \texttt{PrHybridSeeding}~\cite{HybridSeeding}.
It relies on an iterative approach with three iterations, where high-momentum tracks are reconstructed first, allowing to progressively clean up the environment for further iterations with enlarged tolerance windows to search for lower momentum track candidates. 
Unlike \hlttwo, the \hltone \HS runs only two independent iterations and removes repeated tracks at the end of the algorithm. 
This choice is driven by the stringent throughput requirements of HLT1.

\subsection{General structure of the algorithm}
The \scifi detector provides the hit position ($x$)
with a high resolution on the bending plane ($x$-$z$) of LHCb thanks to its six vertically positioned $x$-layers. The position of the fibres along the beam axis ($z$) is also precisely known\footnote{Up to few tens of microns after alignment.}, therefore the pattern recognition begins with a search for tracklet candidates in the \xz plane. This first step constitutes the \textit{XZ seeding}, the most computationally expensive step in the reconstruction sequence. Due to the non-zero magnetic field in the SciFi detector, tracks in the \xz plane are described by
\begin{center}
    \begin{equation}
        x(z) = a_x + b_x\zb{} + c_x\zb{}^2(1+\dratio\zb{}),
        \label{eq:trackequation}
    \end{equation}
\end{center}

\noindent
where $\zb{} = z - z_\text{ref}$, $a_x,\ b_x,\ c_x$ are terms of the parabola that need to be determined, and \dratio, which describes the decreasing magnetic field along the $z$ direction, is fixed from simulation. The reference point $z_\text{ref}=8525\mm$ is chosen as a mid-point in the SciFi to improve numerical stability.

The track model is then fitted to the \xz tracklets and a clone\footnote{Clones are tracks that share more than 50\% of their hits.} removal is performed to suppress the large fraction of duplicates. The surviving candidates are then extended to the remaining six $u$-$v$ layers, where the track description along the non-bending ($y$-$z$) plane is also determined, in the \textit{UV seeding} step. The various steps are described below, with a particular focus on new features of the adaption of the algorithm to HLT1.

\subsection{XZ seeding}

\begin{figure}[h]
    \centering
    \includegraphics[width=0.5\textwidth]{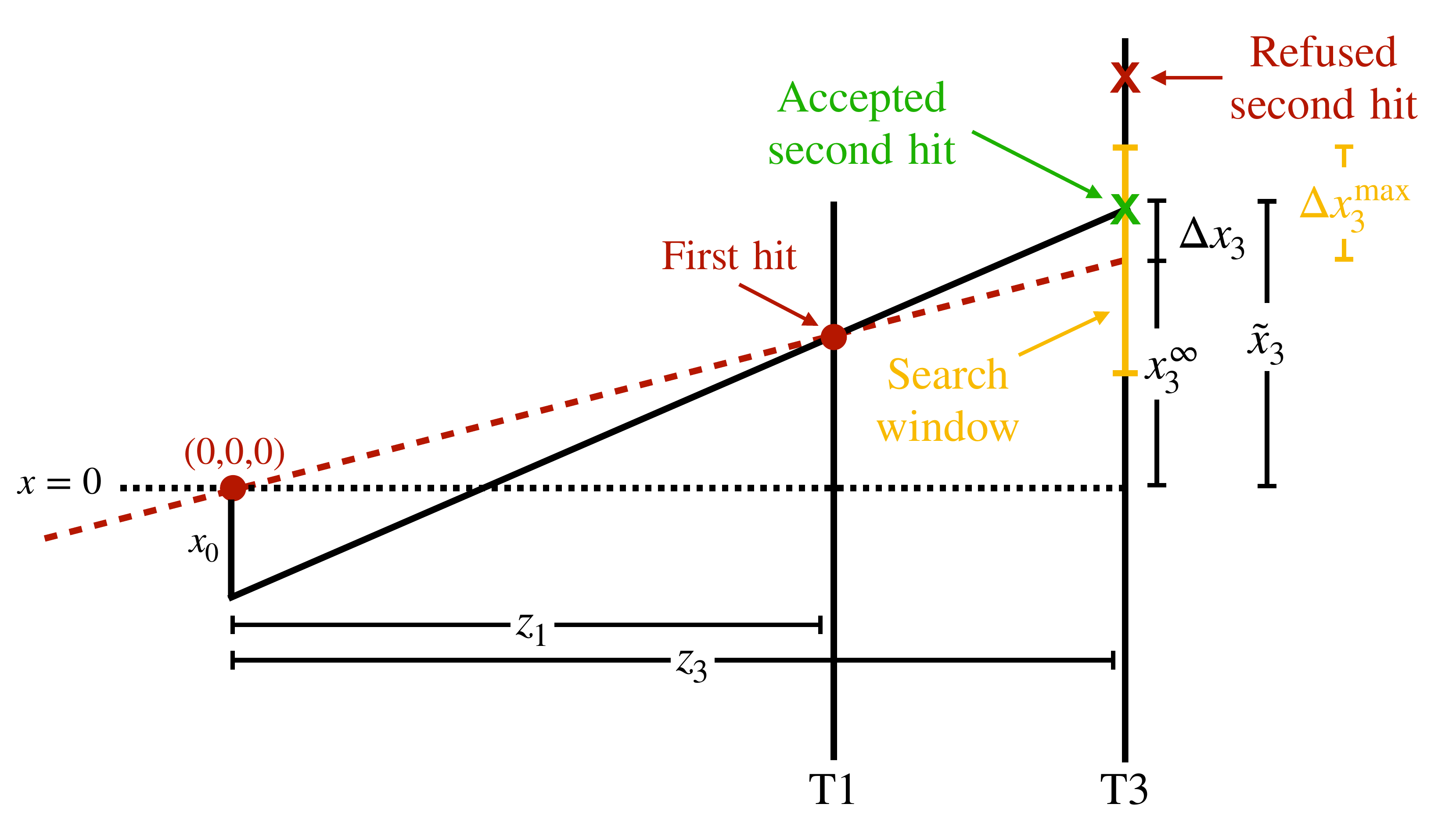}
    \caption{Schematic of the construction of two-hit combinations in the $xz$-plane.}
    \label{fig:SeedingXZ_twohit}
\end{figure}

\begin{figure}[h]
    \centering
    \includegraphics[width=0.5\textwidth]{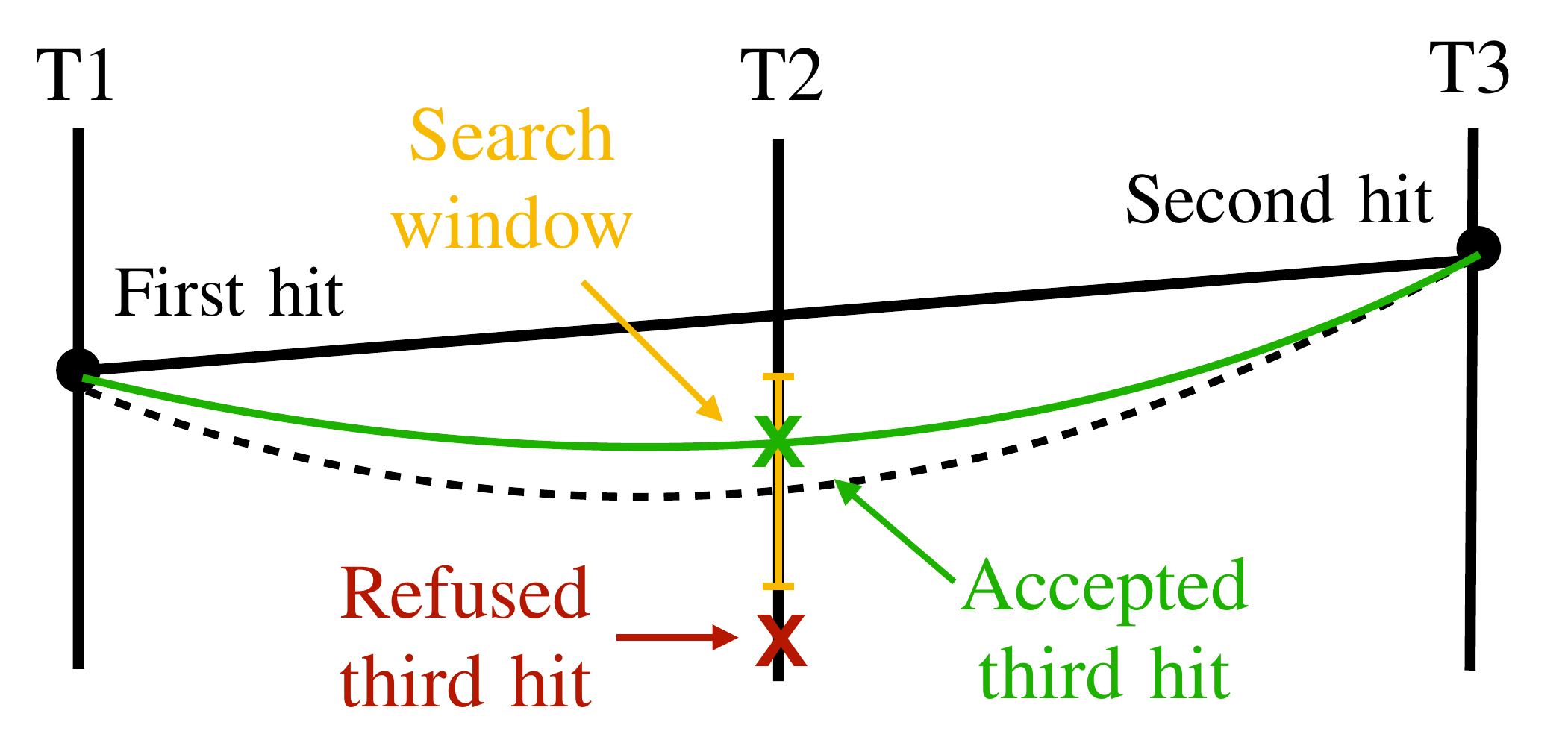}
    \caption{Schematic of the construction of three-hit combinations in the $xz$-plane.}
    \label{fig:SeedingXZ_threehit}
\end{figure}

The seeding in the \xz plane consists of a sequence of steps to reconstruct \xz tracklets that can be run in several iterations as described before. 
Each iteration of the XZ seeding starts from a set of three initial $x$-layers, one in each of the SciFi stations, to build first two- then three-hit combinations. Hits from the remaining $x$-layers are then added to complete the $x$-$z$ seeding, before a track fit in the \xz plane and a removal of clone tracks is performed. 
To speed up the algorithm for HLT1, only two iterations are run, both targeting particles of minimum momentum $p_\mathrm{min}>\SI{3}{GeV/c}$, but with disjoint sets of initial $x$-layers to accommodate hit detection inefficiencies.
The construction of two- and three-hit combinations in the XZ seeding are depicted in Fig. \ref{fig:SeedingXZ_twohit} and Fig. \ref{fig:SeedingXZ_threehit}, respectively.

All hits from the initial T1 $x$-layer are combined with all hits from the initial T3 $x$-layer that lie in a predefined search window to form two-hit combinations. The position of the window is defined by a straight line extrapolation from the T1 hit to T3 assuming the track originated from the centre of the collision region ($x=y=z=0$) and its size is given by the initial track momentum assumption of $p_\mathrm{min}>\SI{3}{GeV/c}$ for which the seeding is tuned, as explained in the next section. 
The two-hit combinations are back-propagated as straight lines to $z=0$ and an estimate for the track momentum is derived from $x_0=x(z=0)$. The momentum is then used to define a search window in the initial T2 $x$-layer to look for the third hits and build three-hit combinations compatible with the track model in Eq.~\ref{eq:trackequation}. The track model of the three-hit combination is then extended to the three remaining $x$-layers and hits within tolerance windows of 1\mm are added. Only \xz tracklets with at least 5 hits from all $x$-layers are kept. Finally, a least-squares fit to the \xz track model in Eq.~\ref{eq:trackequation} is performed, with $a_x$, $b_x$ and $c_x$ as free parameters and \dratio fixed, and track candidates with bad $\chi^2$ are filtered out.

\subsection{Momentum estimation}
The XZ seeding relies on tolerance windows to minimise complexity while maintaining high efficiencies. These tolerance windows are tuned using a sample of simulated $B_s^0 \to \phi \phi$ decays (with 
$\phi\to K^+K^-$)
and the window sizes are parameterised as a function of: momentum, the choice of the initial $x$-layers and other topological variables. This ensures that the maximum number of hits is contained in the window. 

\subsubsection{Inverse momentum fit}
\label{sec:invP}
Long and Downstream tracks can be considered to a good approximation, as coming from the origin at $(0,0,0)$.
A track of infinite momentum coming from that point and passing through a first hit of coordinates $(x_1,z_1)$ in T1 will pass the $z=z_3$ detection plane in T3 at
\begin{equation}
    x_3^{\infty} = x_1 \times \frac{z_3}{z_1}.
\end{equation}

\begin{figure}[h]
    \centering
    \includegraphics[width=0.4\textwidth]{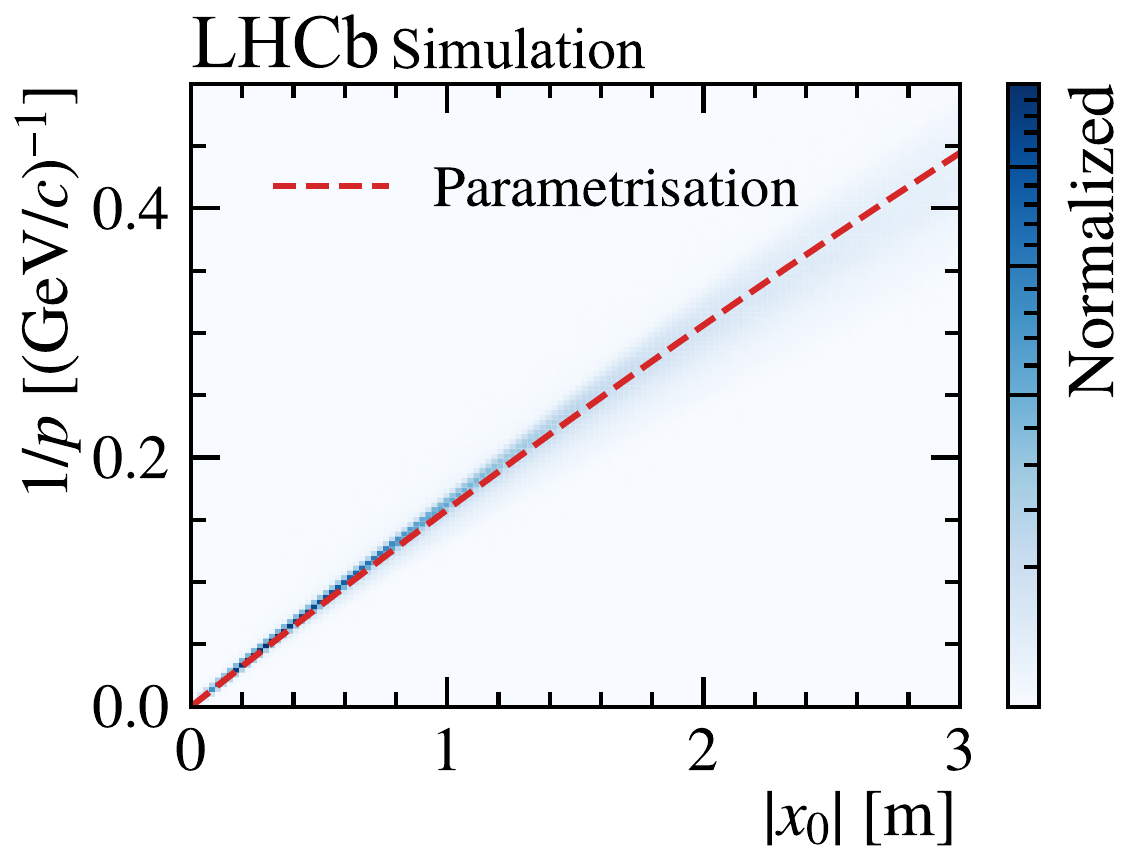}
    \caption{Inverse momentum ($1/p$) distribution as a function of $\lvert x_0 \rvert$ for simulated tracks, where $x_0$ is the $x$-position of the two-hit combination back-propagated to $z=0$. The fitted parametrisation is overlaid.}
    \label{fig:invPFits}
\end{figure}

Opening a tolerance for finite momenta in the search for the second hit is then equivalent to opening a tolerance around this $x_3^\infty$ position.
A useful variable that contains information about the integrated deflection of the track in the magnet is the back-propagated $x$ of a pair of hits, $x_0$, defined as
\begin{equation}
    x_0 = x_1 - z_1 \times t_{13}.
\end{equation}
with
    \begin{equation}
        t_{13} = \frac{\left(x_3-x_1\right)}{\left(z_3-z_1\right)}
    \end{equation}
being the slope between the two first hits.
Figure~\ref{fig:invPFits} shows the relation between this $x_0$ variable and the inverse momentum for Long tracks for simulated events. The dependency of the inverse momentum is thus parameterised as
\begin{align}
\label{eq:invp}
\frac{1}{p} = f(x_0) = a \vert x_0\vert + b x_0^2
\end{align}
and the fitted parameters are
\begin{equation}
    \begin{aligned}
        a&=1.6322\times10^{-7}\: (\gevc)^{-1} \mm^{-1}, \\
        b&=-5.0217\times10^{-12}\: (\gevc)^{-1} \mm^{-2}.
    \end{aligned}
\end{equation}

 These parameters depend on the choice of initial layers, since the integrated field between the two measurement points changes with the layer positions. However, the variation is small enough that a single set of parameters is used for both iterations.

\subsubsection{Second hit tolerance window}
The difference, $\Delta x_3$, between the measured $x$-position $\tilde{x}_3$ of a hit in T3 and the predicted position $x_3^\infty$ in T3, assuming the track originated from $(0,0,0)$ and has infinite momentum, is proportional to $x_0$, following
\begin{equation}
\label{eq:delta3}
    \Delta x_3 = \frac{z_3-z_1}{z_1}x_0.
\end{equation}

A momentum criterion can be translated into a cut on $x_0$ through the model of Eq.~\ref{eq:invp},  and into a tolerance on $\Delta x_3$ using Eq.~\ref{eq:delta3}. The following dependency of the $\Delta x_3$ tolerance on $p_{\rm min}$ is obtained as
\begin{align}
\label{eq:tolHp}
{\Delta x_3}^\text{max} = \frac{z_3-z_1}{2 z_1 b}\biggl(-a + \sqrt{a^2 + \frac{4 b}{p_\text{min}}}\biggr),
\end{align}
which is independent from the choice of the T1 and T3 layers.

The tolerance $\pm\Delta x_3^\text{max}$ is symmetric around $x_3^\infty$, but the actual distribution of hits from finite-momentum tracks is not. For a given first-hit position $x_1$, the population of positive and negative charges arriving at that position is not symmetric: the magnetic field between the VELO and the SciFi preferentially deflects one charge sign towards a given $x_1$ region. As a result, the $\Delta x_3$ distribution at a given $t_x^\infty = x_1/z_1$ is shifted with respect to zero. This effect was also considered in the \hlttwo \texttt{PrHybridSeeding} algorithm.
To minimise the window size needed to achieve a given coverage, the window centre is shifted with the parametrisation
\begin{align}
\label{eq:alphacorr}
        \alphacorr = \frac{\alpha_0}{k ( p_\mathrm{min}-\alpha_1)},
\end{align}
where $k = \frac{z_1}{z_3-z_1}$ and $\alpha_0$ and $\alpha_1$ are obtained from a fit to simulated data. The results are
\begin{equation}
    \begin{aligned}
        \alpha_0 &= 2.180\times10^{6}\: \mevc, \\
        \alpha_1 &= -1073 \: \mevc.
    \end{aligned}
\end{equation}

\begin{figure}[h]
    \centering
 \includegraphics[width=0.48\textwidth]{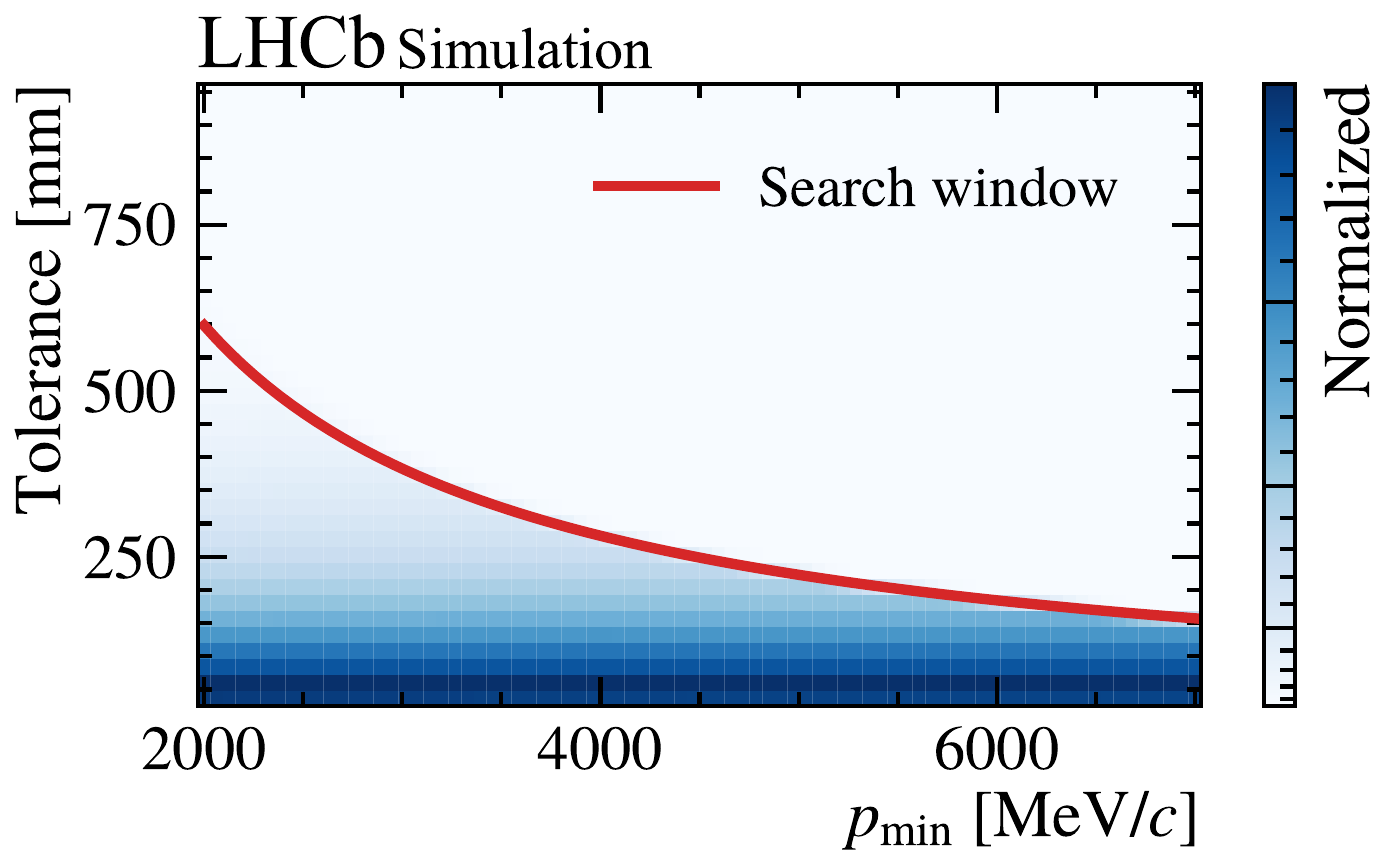}
\caption{Distribution of the  tolerance ($\Delta x_{3}$) required to contain 2-hit combinations as function of $p_{min}$, after applying the charge-asymmetry correction. The parametrised model (in red) contains more than $99{\%}$ of the signal hits across the full momentum range.
    }
    \label{fig:2hit_tolerance}
\end{figure}
The upper and lower limits for the tolerance window are then calculated as
\begin{align}
    x_\text{tol}^\text{lower/upper} = \frac{z_3+\alpha_\text{corr}}{z_1}x_1\mp{\Delta x_3}^\text{max}.
\end{align}
The correction shifts the window center independently of the magnet polarity taking into account the preference in deflection by the charged tracks. Figure~\ref{fig:2hit_tolerance} shows the tolerance required to contain each hit as a function of $p_{\text{min}}$, after applying the correction. The model described by Eq.~\ref{eq:tolHp} is superimposed in red, and shows that the two-hit search window contains more than $99{\%}$ of the signal hits.  

\subsubsection{Third hit tolerance window}

For each two-hit combination, a third hit is searched for in the 2nd layer (T2), as illustrated in Fig.~\ref{fig:SeedingXZ_threehit}. The predicted position in this layer is obtained by extrapolating the slope $t_{13}$ from the first hit:
\begin{center}
    \begin{equation}
        \Delta x_{2} = x_{2} - \left(x_1 + \left(z_2-z_1\right)t_{13} \right),
    \end{equation}
\end{center}
The deviation of the third hit from its prediction measures the track curvature due to the residual magnetic field inside the SciFi. This quantity is proportional to the curvature parameter $c_x$:
\begin{center}
    \begin{equation}
        \Delta x_{2} = k_3 \cdot c_x,
    \end{equation}
\end{center}

where $k_3$ is a geometric factor depending only on the $z$ positions of the three layers and the reference point, computed from the track model including the $d_\text{ratio}$ correction.

\begin{center}
    \begin{equation}
    k_3 = \big[\left(z_{2}^2-z_{1}^2\right)\frac{z_2-z_1}{z_3-z_1}+\left(z_{1}^2-z_{3}^2\right)\big]^{-1}.
    \end{equation}
\end{center}

\begin{figure}[h]
    \centering
    \includegraphics[width=0.48\textwidth]{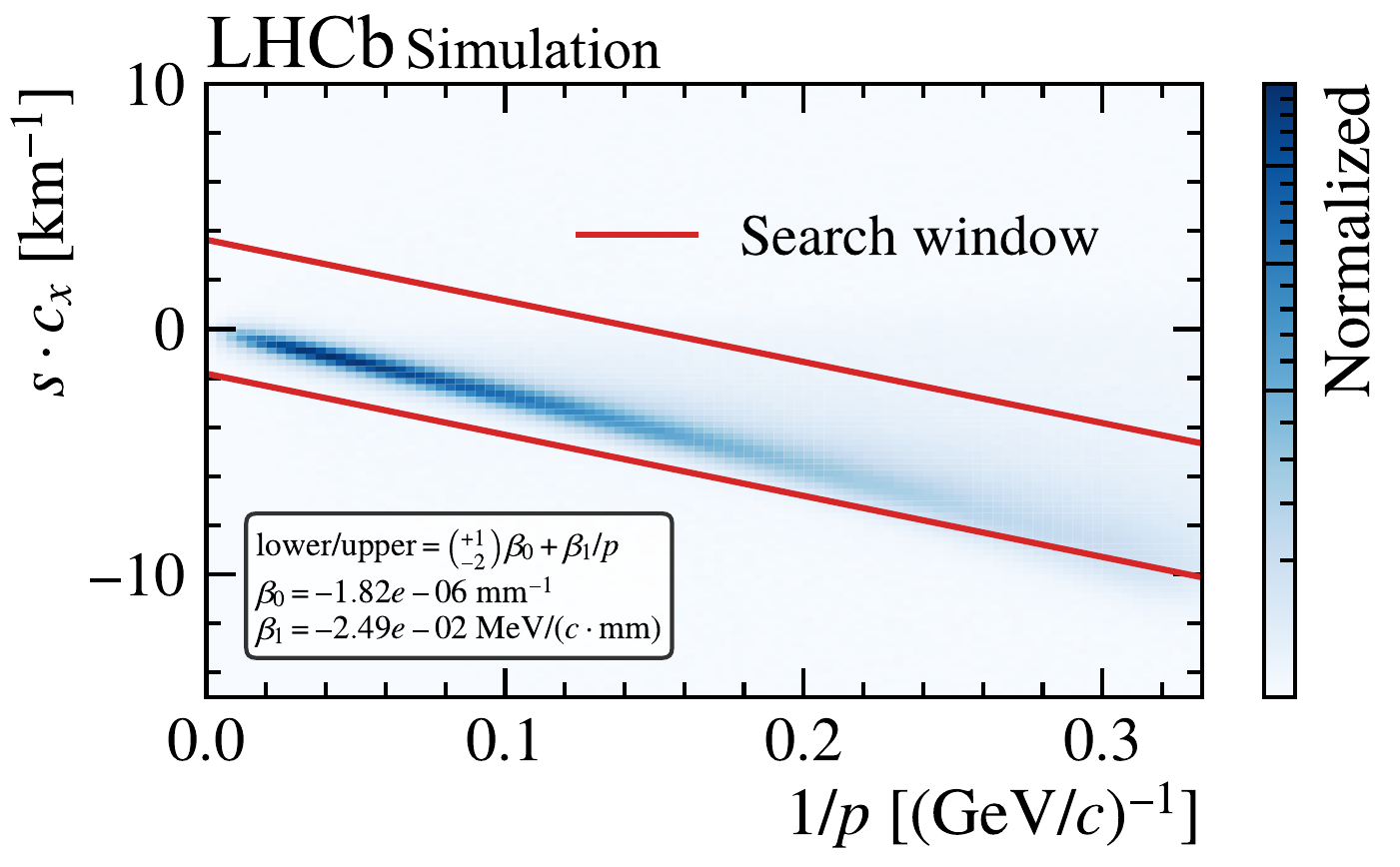}
\caption{Distribution of $s \cdot c_x$ as a function of $1/p$ for tracks with $p > 3 \gevc$ from simulation, where $c_x$ is the curvature parameter of the parabolic track model in Eq.~\ref{eq:trackequation} and $s = \mathrm{sign}(x_0)$. The red lines define the tolerance window, with the parametrisations shown on the plot.}
\label{fig:3hit_tolerance}
\end{figure}

The curvature parameter $c_x$ arises from the residual magnetic field inside the SciFi and is therefore proportional to $q/p$. To visualise this relation independently of the LHCb magnet polarity, Fig.~\ref{fig:3hit_tolerance} shows the distribution of $s \cdot c_x$ as a function of $1/p$, where $s = \text{sign}(x_0)$ cancels the polarity dependence. The main bulk of the distribution follows a linear relation, but the spread is not symmetric: one side exhibits a longer tail, attributed to energy loss or different integrated fields for tracks passing through the outer regions of the magnet.

While the two-hit tolerance effectively measures the track momentum under the assumption that the track originates from $(0, 0, 0)$, the third-hit tolerance measures the momentum using only the residual field in the SciFi. The former has a resolution of approximately 1\% for tracks from the interaction region, while the latter has a resolution of approximately 20\% but is valid for all track types, including downstream tracks that do not originate from the primary vertex.

\subsubsection{Clone killing}
\label{sec:clonekilling}

The two iterations of the algorithm, combined with the multiple combinatorial paths in the track building, produce a large number of clone tracks, with approximately 50\% of the candidates being duplicates. The clone-killing algorithms used in HLT2 compare all pairs of tracks ($O(n^2)$ complexity), which is too slow to be executed on GPUs. A new clone removal algorithm with $O(n)$ complexity has therefore been developed.

For each track, a score is defined as
\begin{equation}
    \sigma = \begin{cases}
        1000 \cdot \chi^2 + c_x \,, & \text{if } n_\text{hits} = 6 \,, \\
        10 \cdot (1000\,\chi^2 + c_x) \,, & \text{if } n_\text{hits} = 5 \,,
    \end{cases}
    \label{eq:seed_clone_score}
\end{equation}

where $\chi^2$ is the fit quality of the $x$--$z$ projection and $c_x$ is the curvature parameter. The score favours tracks with more hits and better fit quality. The inclusion of $c_x$ provides a deterministic tie-breaker when two tracks have identical $\chi^2$ values, preferring higher-momentum tracks.
The factor of 10 strongly favours six-hit tracks: a five-hit track is only preferred when its $\chi^2$ is more than ten times smaller than that of an overlapping six-hit track.

The algorithm operates on a shared lookup table indexed by SciFi hit, with one entry per hit across all layers. It proceeds in two passes, both fully parallelised on the GPU.

In the first pass, each track writes its score to the table entries corresponding to its hits. If multiple tracks share a hit, only the best (lowest) score is retained, using the CUDA \texttt{AtomicMin} operation to guarantee thread-safe updates. After this pass, each hit in the table is associated with the score of the best track that contains it.

In the second pass, each track reads back the scores stored at its hit positions and counts how many of its hits are still associated with its own score. A track is retained if at least three of its hits carry its score, or if at least two do and the track has six hits in total. Tracks that fail this criterion are discarded: their hits have been claimed by better candidates.

This voting-based approach avoids the explicit pairwise comparison of tracks and scales linearly with the number of tracks, making it well suited for GPU execution. It reduces the fraction of clones from about 50\% to about 1--2\%.

\subsection{UV seeding}

The $x$--$z$ track candidates surviving the previous steps contain only hits from the $x$-layers and have no information about the $y$ position of the track. The next step is to add hits from the six stereo layers to recover the $y$--$z$ projection.
Two implementations of the UV seeding have been developed for HLT1: an iterative algorithm over all stereo hits in an initial layer within the selection window, and 
one algorithm using bitwise Hough clusters, the latter following the \hlttwo \texttt{PrHybridSeeding} algorithm strategy. Both approaches have similar tracking performance and the first is taken by default since it was first implemented and validated with real data.   

 For a given $x$--$z$ projection, the predicted $x$ position at the $z$ coordinate of a stereo layer is known from the fitted track parameters. The measured coordinate $x_\text{hit}$ in the local frame of the stereo layer is related to the $y$ position by
\begin{center}
    \begin{equation}
        y_0 = \frac{x(z_0)-x_0}{\mathrm{tan}\,\alpha}.
    \end{equation}
\end{center}
$\alpha $ being the stereo angle of the layer ($\pm 5^\circ$).
In the SciFi region, the magnetic field is predominantly oriented along the $y$ direction, and trajectories in $y$-$z$ plane are therefore expected to be approximately straight lines.

In the iterative procedure for each $x$--$z$ projection, all stereo hits within the selection window are collected in a first layer. For each such hit, a $t_y$ estimate is computed and used to predict the expected $x_\text{hit}$ position in the remaining five stereo layers. The closest hit within a narrow tolerance of $2\,\text{mm}$ around the predicted position is selected in each layer. When a hit is found, the $t_y$ estimate is refined by averaging the current value with the new measurement, progressively improving the prediction for subsequent layers. At most one hit combination is constructed per initial hit. All combinations are fitted with a least-squares fit to the $y$--$z$ straight-line model, and the candidate with the largest number of hits is selected. Among candidates with the same number of hits, the one with the best fit quality is preferred. Similarly to the XZ seeding, a second pass using a different initial layer is performed to cover possible hit inefficiencies.

In the Hough-cluster approach, instead of iterating over hits in a single initial layer, a one-dimensional Hough transform is performed over all stereo layers. Each candidate hit is mapped to a $t_y$ bin, and the most populated bins are identified. Up to eight of the best bins are selected and each is subdivided into several finer bins to provide a more accurate position. For each refined $t_y$ hypothesis, the closest hit within the same $2\,\text{mm}$ tolerance is attached in each layer. The resulting combinations are fitted and the best candidate is selected using the same criteria as the default approach.

The UV seeding recovers the three-dimensional track information, and serves to reject fake $x$--$z$ projections. After this process the T-track reconstruction is complete and they are made available to other algorithms.

\section{Matching algorithm}
\label{sec:matching}

The HLT1 \matching algorithm is inspired by the HLT2 \texttt{PrMatchNN} algorithm, which is one of the two complementary Long track reconstruction methods used in LHCb. The principle of this algorithm is as follows. First, track segments are reconstructed independently in the VELO and SciFi subdetectors using seeding algorithms. These segments are then linked together in the \matching step to form Long tracks, and the best UT hits are added if the UT is included in the data taking and they are found to be compatible with the trajectory. The UT addition reduces fake-track reconstruction and improves momentum resolution.
The same principle is retained in the HLT1 implementation of this algorithm, using the VELO \texttt{Search by Triplet} algorithm \cite{Campora:search_by_triplet} and the \HS algorithm previously described, respectively.

\begin{figure}
    \centering
    \includegraphics[width=0.5\textwidth]{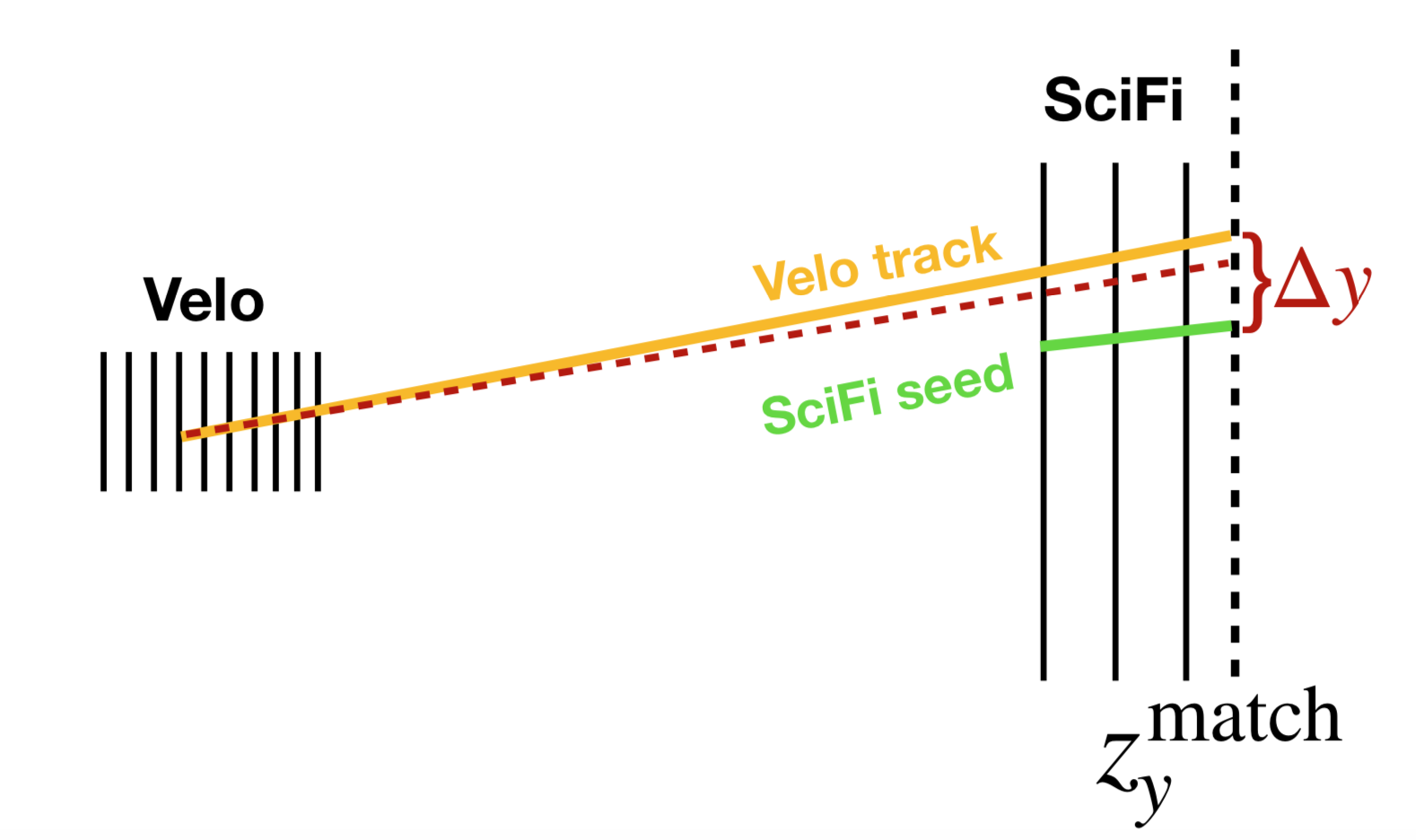}\hfill \includegraphics[width=0.5\textwidth]{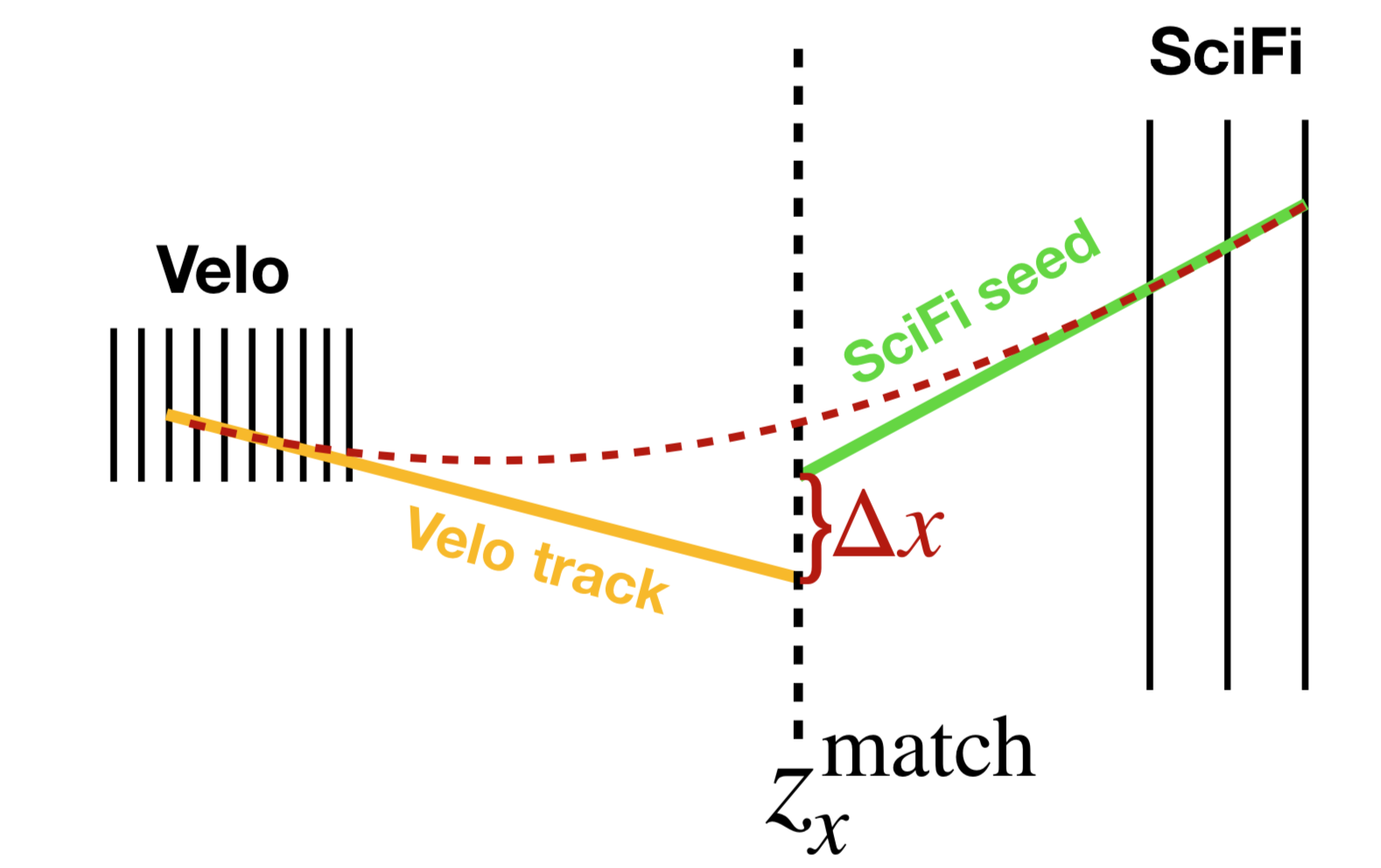}
    \caption{Schematic of the \textit{VeloSciFi} matching in the (left) $y$ and (right) $x$ direction.}
    \label{fig:skmatching}
\end{figure}

Considering that the magnetic field in LHCb is mostly along the $y$ direction and charged particle trajectories show negligible bending in the $y$-$z$ plane, flight paths are expected to be straight lines to a good approximation. In order to match track segments from VELO and SciFi in the $y$ dimension, both segments are extrapolated as straight lines up to $\zmatchy = \SI{10}{\meter}$ as illustrated in Fig.~\ref{fig:skmatching}.
In order to integrate the small $y-$bending due to a residual $B_z$ component of the magnetic field, tracks are matched in $y$ at the end of the SciFi detector.
A correction to account for this residual bending effect has been implemented in the $y$-$z$ track model of HLT2, but this was found to have negligible impact on the HLT1 tracking efficiency and fake rate, and the simpler straight-line model was chosen instead. 
\subsection{Magnetic field parameterisation}

The effect of the magnetic field between VELO and SciFi needs to be taken into account in order to correctly reconstruct charged particles. As already mentioned, LHCb uses a dipole magnet with a vertical field, therefore tracks are expected to mostly bend along the $x$ direction. The most accurate way to predict the particle trajectories in the tracking stations would be to use the full LHCb magnetic field map. However, performing the proper extrapolation with full LHCb magnetic field map at HLT1 requires a large amount of memory access and expensive computations. Given that the processing speed is essential in HLT1, it has been decided to parameterise the bending of tracks in the $x$ direction.

For long-distance extrapolation, the effect of the magnetic field can be viewed as a ``kink'' of the VELO track in a certain position inside the magnet, \zmatchx. Both VELO and SciFi track segments are extrapolated to this position following:

\begin{equation}
    \begin{aligned}
     x_{\rm VELO}^{z_{\rm match}} & = x_{\rm VELO}^{\zref} + (\zmatchx - \zref) \cdot t_x^{\rm VELO}
    \\
     x_{\rm SciFi}^{z_{\rm match}} & = x_{\rm SciFi}^{\zref} + (\zmatchx - \zref) \cdot t_x^{\rm SciFi} \\
    \end{aligned}
\end{equation}

where \zref is the reference position, corresponding to the last VELO station or SciFi station for the VELO and SciFi segment, respectively. Depending on the track momentum, the optimal position to match the two track segments will differ for each pair of segments. The $z$-position of the kink, \zmatchx, is therefore not a fixed value but encapsulates the effect of the magnetic field:

\begin{equation}
\begin{aligned}
    \zmatchx
        & =  A_x + B_x \times \lvert\Delta t_x\rvert  + C_x \times \lvert\Delta t_x\rvert^2  \\
        & + D_x \times \lvert x_{\rm SciFi}^{\zref}\rvert   + E_x \times {t_x^{\rm VELO}}^2
\end{aligned}
\label{eq:z_match}
\end{equation}

As shown in Eq.~\ref{eq:z_match}, at first order, the parameterisation depends on the difference in slopes of the two track segments, $\lvert\Delta t_x\rvert$, which is in turn inversely proportional to the track momentum. The additional terms account for the residual field outside the magnet. The parameters $A_x$, $B_x$, $C_x$, $D_x$ and $E_x$ are calculated from simulation: hits from each true simulated particle are used to reconstruct the true VELO and SciFi straight-line segments, then their intersection point is used to fit the model in Eq.~\ref{eq:z_match}. Results are shown in Table~\ref{tab:mag_params}.

\begin{table}[h]
    \centering
    \caption{Values for the parameters of the magnetic field parameterisation obtained from simulation.}
    \begin{tabular}{c r}
    \toprule
    Parameter & Value \\
    \midrule
     $A_x$  & \SI{5287.6}{\milli\meter} \\
     $B_x$  & \SI{-7.98878}{\milli\meter} \\
     $C_x$  & \SI{317.683}{\milli\meter} \\
     $D_x$  & 0.0119379 \\
     $E_x$  & \SI{-1418.42}{\milli\meter} \\
     $A_{\gamma}$ & $7.707063 \times 10^{-11}$ \\
     $B_{\gamma}$ & $-7.817364 \times 10^{-3} \ \rm{c/MeV}$  \\
    \bottomrule
    \end{tabular}
    \label{tab:mag_params}
\end{table}

For short distance extrapolation, such as extrapolating VELO segment to the UT, a uniform constant magnetic field in the $y$ direction is assumed. The trajectory can thus be described as

\begin{equation}
    \begin{aligned}
    x(z) &= x_{\rm VELO} + t_x^{\rm VELO} \times (z - \zref) + \gamma_{x} \times (z - \zref)^{2} \\
    y(z) &= y_{\rm VELO} + t_y^{\rm VELO} \times (z - \zref) \\
    \end{aligned}
\label{eq:velo_ut_extrapolation}
\end{equation}

where $\gamma_{x}$ is the parameter that depends on the momentum of the track and the integrated magnetic field between VELO and UT, which can be parameterised as
\begin{equation}
    \gamma_{x} = A_{\gamma} + B_{\gamma} \times \frac{q}{p} \\
\label{eq:gamma_parametrisation}
\end{equation}
The parameters $A_{\gamma}$ and $B_{\gamma}$ are extracted from simulation and are shown in Table \ref{tab:mag_params}.

\subsection{Candidate preselections}

Since the \texttt{Matching} algorithm must iterate over all possible VELO–SciFi combinations, a candidate preselection is applied to reduce processing time. Only combinations that satisfy all of the following conditions are considered:

\begin{equation}
    \begin{aligned}
        &\lvert \Delta x\rvert < \SI{20}{\milli\meter},
        &\ 
        &\lvert \Delta y\rvert < \SI{150}{\milli\meter},
        \\
        &\lvert \Delta t_x\rvert < 1.5,
        &\ 
        &\lvert \Delta t_y\rvert < 0.02,
    \end{aligned}
\end{equation}
\\
where $\lvert \Delta x \rvert$ and $\lvert \Delta y \rvert$ are the differences in extrapolated positions at the matching planes $\zmatchx$ and $\zmatchy$, respectively, and $\lvert \Delta t_x \rvert$ and $\lvert \Delta t_y \rvert$ are the differences in their slopes. These cuts are fiducial cuts obtained from simulation and reject a large number of random combinations while retaining more than $99{\%}$ of true matches. 

After the preselection there are still several surviving candidates per SciFi seed. A match quality variable is constructed to select the candidates that most likely originate from a true particle, using the four matching variables

\begin{equation}
    \chisqmatch = \frac{\Delta x^2}{\tolx} + \frac{\Delta y^2}{\toly} + \frac{\Delta t_x^2}{\toltx} + \frac{\Delta t_y^2}{\tolty}
\end{equation}

where $\tolx, \ \toly, \ \toltx, \ \tolty $ are tolerances taking into account the position and slope resolutions. Treating the track as a straight line and using the known VELO and SciFi track position and slope resolutions, they are reduced to the expressions
\begin{equation}
    \begin{aligned}
        &\tolx  &= &\ \frac{4}{5}[ (\SI{10.12}{\milli\meter})^2 + (\SI{101.0}{\milli\meter})^2 \times \Delta t_x^2 ]
        \\
        &\toly  &= &\ \frac{1}{5}[ (\SI{1.59}{\milli\meter})^2 + (\SI{212.1}{\milli\meter})^2 \times \rho^{\rm VELO} ]
        \\
        &\toltx &= &\ \frac{1}{2}
        \\
        &\tolty &= &\ \frac{1}{937.5}
    \end{aligned}
\end{equation}
\\
where $\rho^{\rm VELO} = {t_x^{\rm VELO}}^2 + {t_y^{\rm VELO}}^2$.

\begin{figure}
    \centering
    \includegraphics[width=0.45\textwidth]{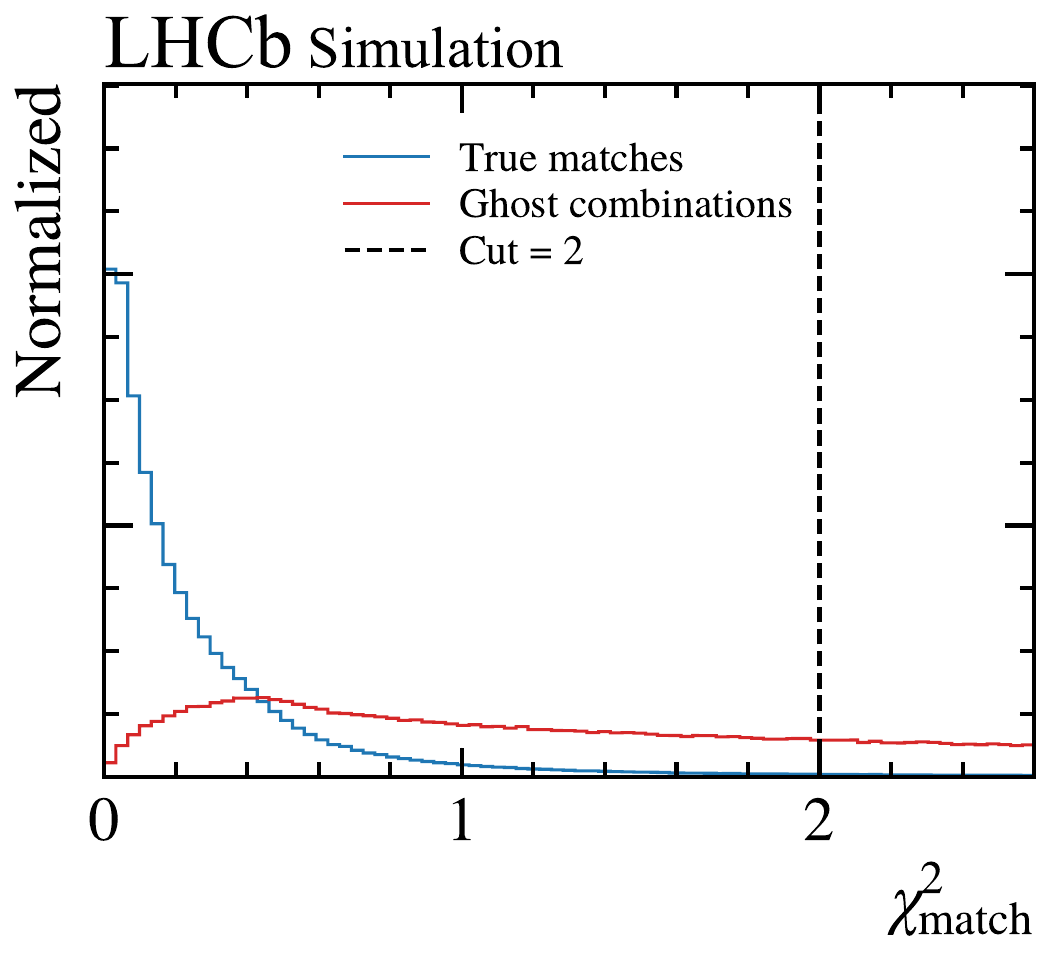}\hfill 
    \caption{Distribution of $\chi^2_\text{match}$ for true matches and ghost combinations after applying the preselection requirements, evaluated on simulated proton-proton collisions with minimal trigger requirements. The vertical dashed line indicates the cut at $\chi^2_\text{match} < 2$.}
    \label{fig:matching}
\end{figure}

The expression of $\toly$ includes only information from the VELO detector, as the VELO state provides a more accurate measurement of the track angle. Figure~\ref{fig:matching} shows the $\chisqmatch$ distribution for fake combinations (ghosts) and true tracks, after applying the fiducial cuts described previously. The criteria $\chisqmatch < 2$ is then applied to remove the bulk of ghost tracks while retaining a high fraction of true matches.

\subsection{Searching for best UT hits}

If UT sub-detector hits are available\footnote{This flexibility was particularly important during the beginning of LHCb Run 3, where the UT sub-detector was in a commissioning phase.}, the algorithm attempts to associate them with each preselected VELO–SciFi pair. The candidate momentum is first used to determine the expected trajectory between VELO and UT using Eqs. \ref{eq:velo_ut_extrapolation} and \ref{eq:gamma_parametrisation}. The VELO segment is then extrapolated sequentially to the UT layers.

At each layer, a loose search window is defined around the extrapolated position. When one or more UT hits are found within this window, their positions are used to refine the track trajectory by updating the parameter $\gamma_x$. The algorithm then proceeds to the remaining UT layers, opening a tighter search window in each. If a UT hit is found within a tight search window, the closest hit to the extrapolated trajectory position is attached to the track.

If no UT hit is found within the loose search window of a given layer, the algorithm continues to the next UT layer and repeats the loose-window search. In the final selection, only candidates with more than two associated UT hits are retained.

The definition of the tight and loose search windows in each UT layer is not constant, but can be parameterized as a function of $1/p$ to account for the effects of multiple scattering, as follows

\begin{equation}
    \begin{aligned}
        \rm{Loose}_{x}^{i}
        &= \min\bigl(
            \max(
                A_L^{i} + \frac{B_L^{i}}{p},
                x_{L}^{\min,i}
            ),
            x_{L}^{\max,i}
        \bigr), 
        \\
        \rm{Tight}_{x}^{i}
        &= \min\bigl(
            \max(
                A_T^{i} + \frac{B_T^{i}}{p},
                x_{T}^{\min,i}
            ),
            x_{T}^{\max,i}
        \bigr), 
        \\
        &\qquad
        i \in \{\rm{UTaX}, \rm{UTaU}, \rm{UTbV}, \rm{UTbX}\}
        \nonumber
    \end{aligned}
\end{equation}

\noindent
where UTaX, UTaU, UTbV, and UTbX are the names of the layers in the UT detector.

The parameters $A_L^{i}$, $B_L^{i}$, $A_T^{i}$, $B_T^{i}$, $x_{L}^{\min,i}$, $x_{L}^{\max,i}$, $x_{T}^{\min,i}$, and $x_{T}^{\max,i}$ are extracted from simulation and are listed in Table~\ref{tbl:ut_search_window_values}.

\begin{table}[h]
    \centering
    \begin{tabular}{c r}
    \toprule
    Parameter & Value \\
    \midrule
     $A_L^{\{\rm{UTaX} \| \rm{UTaU}\}}$  & $0.8333 \: \rm{mm}$ \\[1ex]
     $A_L^{\rm{UTbV}}$  & $1.3333 \: \rm{mm}$ \\[1ex]
     $B_L^{\{\rm{UTaX} \| \rm{UTaU} \| \rm{UTbV}\}}$  & $3.3333 \: \rm{mm\ GeV/c}$ \\[1ex]
     $x_{L}^{\min,\{\rm{UTaX} \| \rm{UTaU}\}}$ & $1.2 \: \rm{mm}$ \\[1ex]
     $x_{L}^{\min,\rm{UTbV}}$ & $1.5 \: \rm{mm}$ \\[1ex]
     $x_{L}^{\max,\{\rm{UTaX} \| \rm{UTaU}\}}$ & $8.5 \: \rm{mm}$ \\[1ex]
     $x_{L}^{\max,\rm{UTbV}}$ & $9.5 \: \rm{mm}$ \\[1ex]
    \midrule
     $A_T^{\rm{UTaU}}$  & $0.2 \: \rm{mm}$ \\[1ex]
     $A_T^{\{\rm{UTbV} \| \rm{UTbX}\}}$  & $0.4 \: \rm{mm}$ \\[1ex]
     $B_T^{\rm{UTaU}}$  & $0.6 \: \rm{mm\ GeV/c}$ \\[1ex]
     $B_T^{\{\rm{UTbV} \| \rm{UTbX}\}}$  & $1.2 \: \rm{mm\ GeV/c}$ \\[1ex]
     $x_{T}^{\min,\rm{UTaU}}$ & $0.5 \: \rm{mm}$ \\[1ex]
     $x_{T}^{\min,\{\rm{UTbV} \| \rm{UTbX}\}}$ & $1.0 \: \rm{mm}$ \\[1ex]
     $x_{T}^{\max,\rm{UTaU}}$ & $2.0 \: \rm{mm}$ \\[1ex]
     $x_{T}^{\max,\{\rm{UTbV} \| \rm{UTbX}\}}$ & $4.0 \: \rm{mm}$ \\[1ex]
    \bottomrule
    \end{tabular}
    \caption{Values of the loose and tight search window parameterizations for UT hit searching, extracted from simulation.}
    \label{tbl:ut_search_window_values}
\end{table}

\subsection{Final selection}
\label{sec:ghost-killing}
\begin{figure}[htb]
    \centering
    \includegraphics[width=0.45\textwidth]{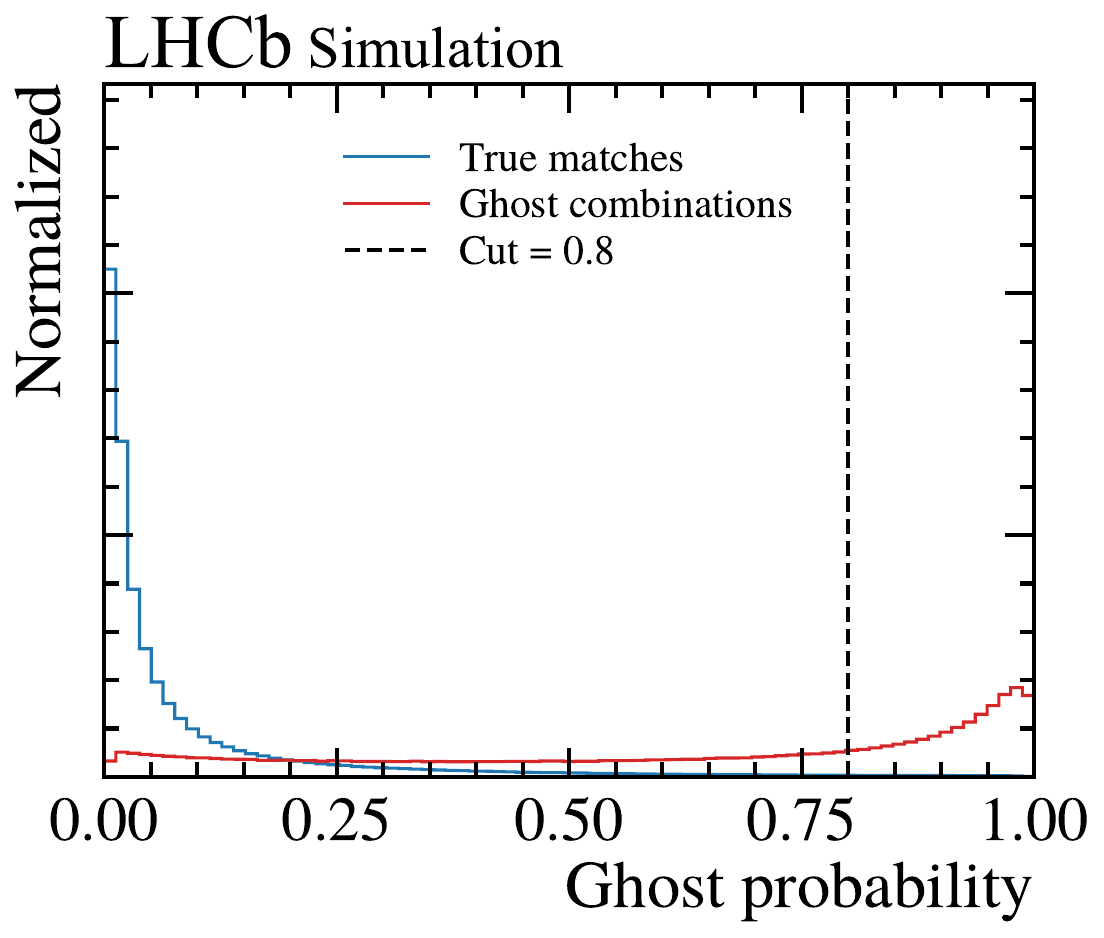}
    \caption{Distribution of the neural network ghost probability for true matches and ghost tracks after the $\chi^2_\text{match} < 2$ requirement, evaluated on simulated minimum bias events. The vertical dashed line indicates the threshold at 0.8.}
    \label{fig:ghost_NN}
\end{figure}

Similarly to the HLT2 implementation, a final ghost-killing neural network is used to perform the final selection. To optimize the throughput of the algorithm, a fully connected neural network (NN) with a single hidden layer of 32 nodes is trained to evaluate the probability that a given candidate corresponds to a fake track. The input variables are $\zmatchx$, $\Delta x$, $\Delta y$, $\Delta t_x$, $\Delta t_y$, $\chi^2_{\rm match}$, $\eta^{\rm VELO}$, and $\rm nHits^{\rm SciFi}$. The output of this neural network is referred to as the ``ghost probability''. 
Figure~\ref{fig:ghost_NN} shows the distribution of the neural network ghost probability for true matches and ghost tracks. 

Only tracks with a ghost probability smaller than 0.8 are retained in the final selection. After this requirement, the ghost probability is further used to remove clone tracks. A clone track is defined as a track that shares either a VELO or a SciFi segment with another track, and whose ghost probability is at least 0.05 higher than that of the corresponding track.
After clone track removal, the remaining tracks constitute the final output of the algorithm.

\section{Graphics Processing Units (GPUs) optimization}
\label{sec:gpus}

The increasing demand for high-throughput computing in high-energy physics is leading experiments to consider alternative computing models.
GPUs, originally developed for image rendering in display applications, are optimised for parallel workloads under the \emph{Single Instruction Multiple Threads} (SIMT) paradigm.
They are therefore well suited to perform the pattern-recognition tasks required to reconstruct events in the high-occupancy environment of LHC experiments.
As discussed in Section~\ref{sec:introduction}, the \runIII \hltone is fully implemented on GPUs.

\subsection{Structure of a GPU}

\begin{figure}
\centering
    \includegraphics[width=0.45\textwidth]{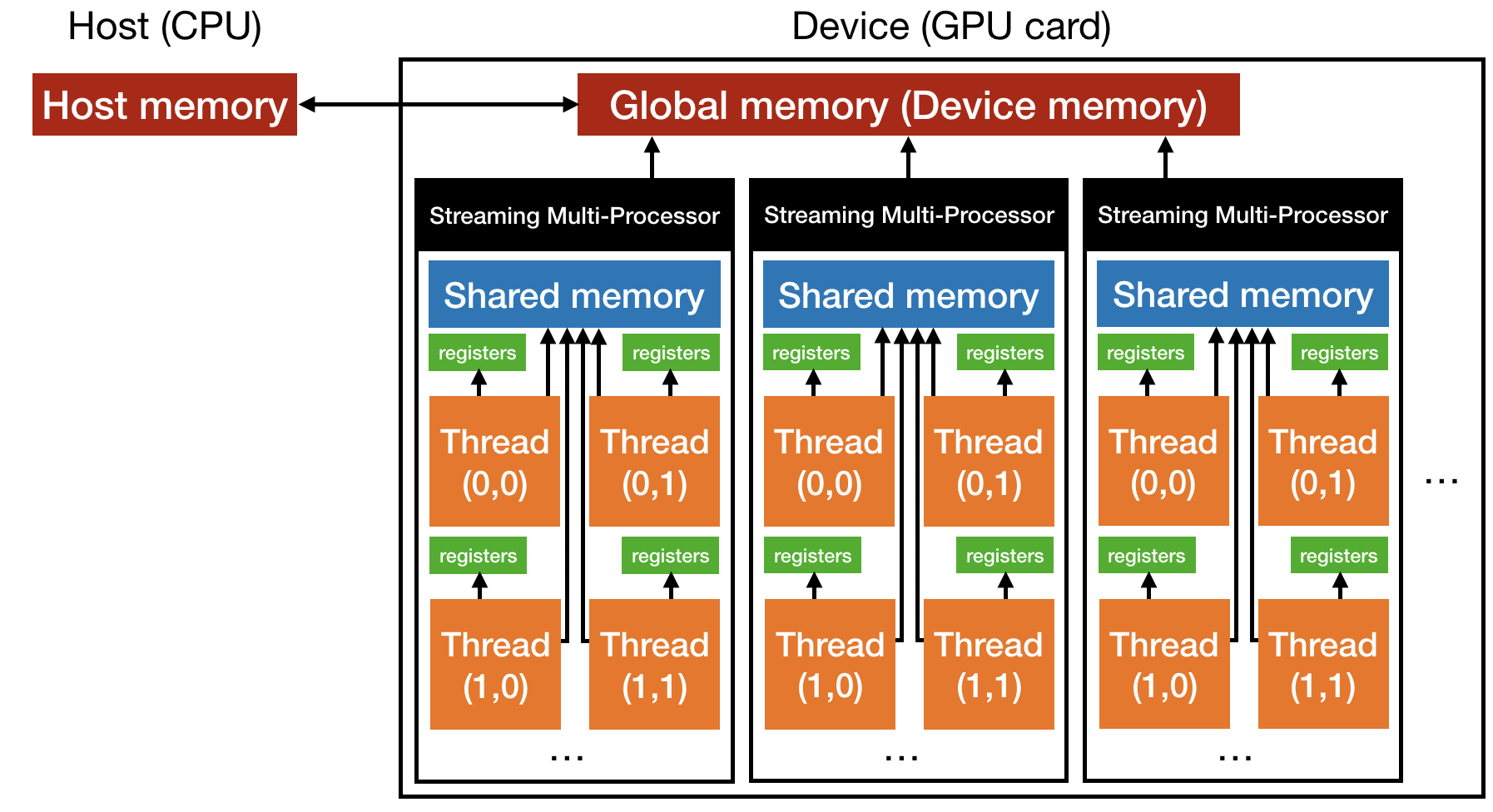}
    \caption{Organisation of threads and memory in a GPU.}
    \label{fig:gpu_structure}
\end{figure}

The key challenges for the GPU implementation of the \HS and  \matching algorithms are memory management and parallelisation.
The structure of a GPU is shown in Fig.~\ref{fig:gpu_structure}.
GPU threads cannot be executed independently, but are instead scheduled and executed in groups of 32 threads, known as warps.
Threads are organised into \emph{Streaming Multiprocessors} (SMs), each of which can execute a fixed number of threads concurrently.
Each SM is equipped with a limited amount of fast \emph{shared memory}, which can be accessed by all threads within the SM.
For parallel execution on the GPU, threads are further organised into blocks, with a configurable number of threads per block.
In addition, each thread has access to a small, private amount of memory, referred to as \emph{registers}, which are the fastest memory available.
Finally, the GPU device provides \emph{global memory}, which is large but relatively slow, shared across all SMs, and used for communication with the host CPU.

\subsection{Memory handling}

Since the different steps of the \hltone \HS algorithm require multiple accesses to hit data, loading this data into \emph{shared memory} and keeping it there across all iterations and processing steps reduces the number of redundant and slow accesses to \emph{global memory}.
Moreover, transferring hit information from \emph{global memory} to \emph{shared memory} can be performed in a highly coalesced manner, allowing the GPU to achieve higher effective memory bandwidth.

The hit data are tiled into chunks corresponding to the upper and lower halves of the detector, and further separated into $x$-layer and $u/v$-layer hits. This split in halves is possible since there is no magnetic field bending in $y$ and only $0.1\%$ of the tracks migrate from one SciFi half to the other. 

In this way, only the data required by each pattern-recognition algorithm are loaded into \emph{shared memory}.
This reduces the amount of \emph{shared memory} required per block, allowing a larger number of blocks to be scheduled on each SM.
Both the XZ and UV seeding stages require approximately \SI{16}{\kilo\byte} of \emph{shared memory} per block on an RTX~A5000 GPU to store the relevant hits of a single event.
This corresponds to four warps, or 128 threads, per block on this device, which is sufficient to efficiently process hits and seed candidates.

Temporary intermediate variables or arrays created during algorithmic computations can be stored in registers.
This is beneficial in terms of performance, as registers are the fastest memory available on a GPU.
The number of registers available per thread is constrained by the number of threads per block, and each algorithm requires a different number of registers.
For example, the Hough-cluster-based version of the UV seeding uses registers to store the Hough histogram instead of relying on global memory.

Reducing unnecessary accesses to \emph{global memory} is crucial for achieving high performance.
For instance, in the XZ seeding algorithm, two-hit combinations are kept in registers while only three-hit combinations are finally stored in global memory.
The intermediate state is kept in registers, avoiding additional global memory traffic.
While this approach increases register usage and may introduce additional warp divergence, it ultimately improves performance by avoiding unnecessary accesses to global memory.

\subsection{Parallelisation}

The optimal number of threads per block depends on both the number of required registers and the amount of required shared memory.
This information is obtained from the CUDA Occupancy Calculator~\cite{cuda-api}, where the number of required registers is taken from CUDA profiling tools, and the amount of required shared memory is determined explicitly by the algorithm.
Both the \HS and \matching algorithms ultimately use 128 threads per block, as this configuration maximises the overall throughput of \hltone on the RTX~A5000.

A further level of parallelism comes from processing many events at the same time.
Each LHCb event is independent and reconstructed with the same sequence of operations, which maps naturally onto the SIMT model.
Since a single event is too small to fully occupy the GPU on its own, events are grouped into batches of around 1000 and processed together.
This event-level parallelism, combined with the thread-level parallelism within each event, allows the GPU resources to be fully utilised.

\section{Performance}
\label{sec:performance}

The performance of the HLT1 \seeding and \matching  algorithms is shown in the following, both in terms of algorithm design and key physics measures. It includes the computing throughput, the track momentum resolution, the physics efficiency, the ability to reject ghost tracks, and the energy consumption of the algorithms.  

\subsection{Throughput}
\begin{figure}[h!]
    \centering
        \includegraphics[width=0.49\textwidth]{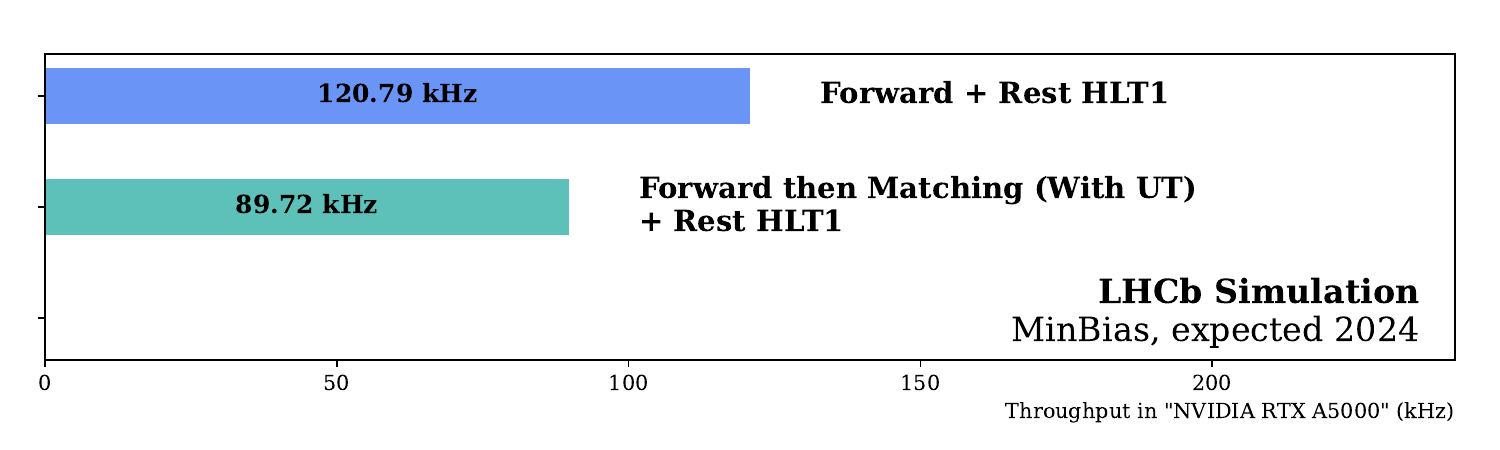}
    \caption{Throughput of the Allen sequence in simulation including the matching algorithm, executed on a NVIDIA RTX A5000 card and compared to other algorithms of the Allen sequence.
    The rest of the algorithms in the sequence refer to the remaining components of HLT1, including the detector decoding, secondary vertex reconstruction, clustering, particle identification and event selection.}
    \label{fig:matching_throughput}
\end{figure}

\begin{figure}[h!]
    \centering
        \includegraphics[width=0.50\textwidth]{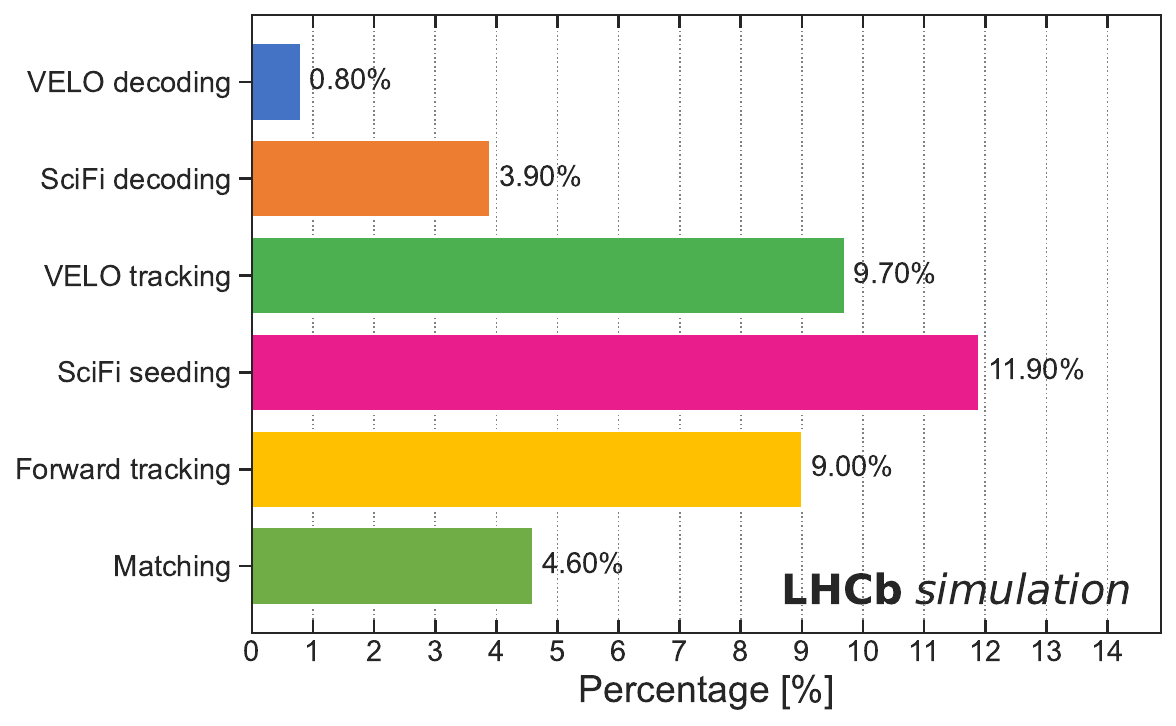}
    \caption{Fraction of the total GPU processing time taken by each algorithm, measured on a NVIDIA RTX~A5000 card. The percentages are computed relative to the full \hltone default sequence used in 2026. }
    \label{fig:matching_algobreakdown}
\end{figure}

The computing performance of the \HS and \matching is measured using simulated minimum bias samples, as they best represent typical inelastic physics collisions with the average occupancy conditions in Run~3. 
Figure~\ref{fig:matching_throughput} shows the throughput of the full \hltone sequence, including the \matching, on a single RTX~A5000 GPU card.
The Run~3 \hltone system is composed of about 500 such cards, so each card must run above 60\,kHz to meet the 30\,MHz requirement.
The measured 89\,kHz per card is well above this value, so the full sequence is sustained at the required rate.
Figure~\ref{fig:matching_algobreakdown} shows the breakdown of \HS and \matching contributions as compared to other algorithms of the sequence.
Within these two algorithms, the \HS takes most of the processing time, since reconstructing track segments from SciFi hits alone requires testing a large number of hit combinations.
The \matching is comparatively fast, as it only has to connect already reconstructed VELO and SciFi segments rather than search for new ones.

\subsection{Momentum resolution}
\begin{figure}[h!]
    \centering
    \includegraphics[width=0.5\textwidth]{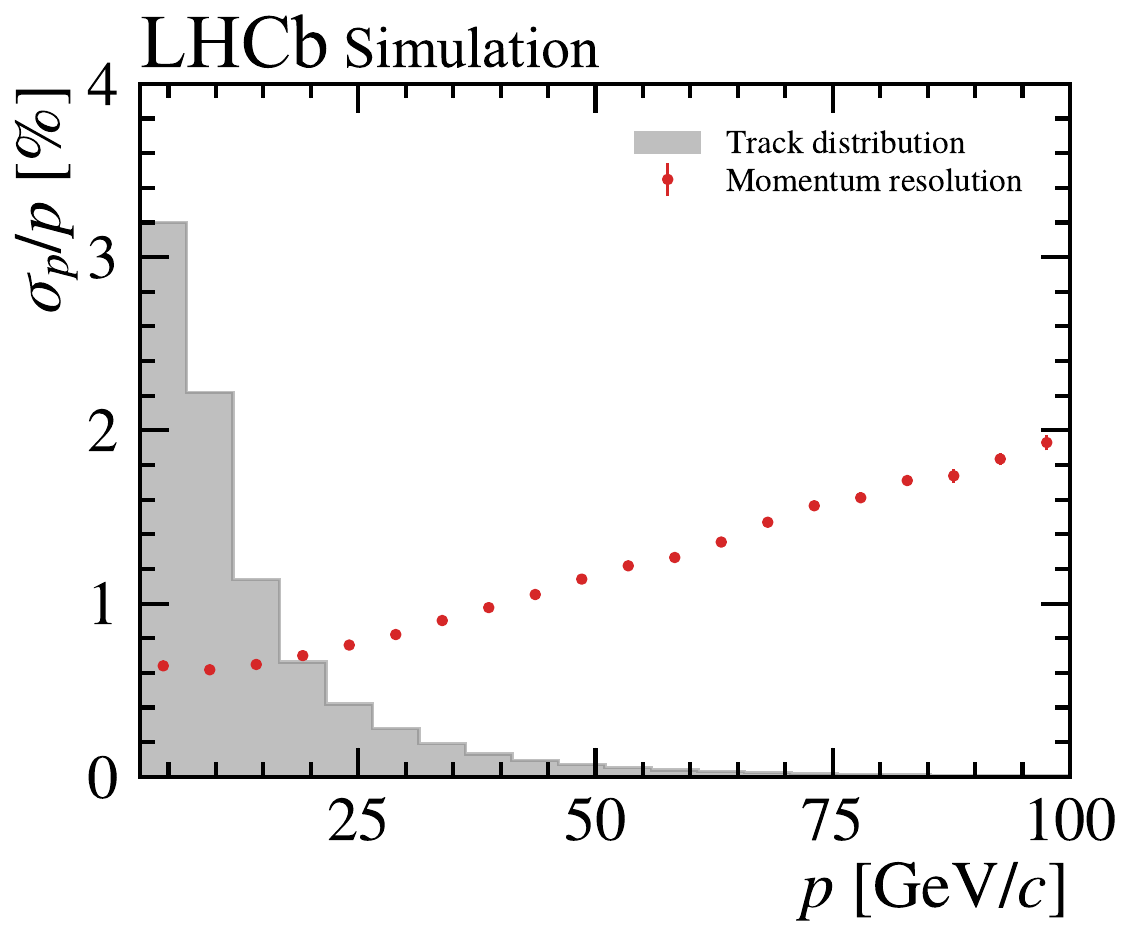}
    \caption{Momentum resolution for HLT1 long tracks from B decays as a function of momentum $p$. The plot shows reconstructed tracks from 10000 simulated $B_s^0 \to \phi \phi$ events where the seeding + matching approach is used for the reconstruction.}
    \label{fig:momentum}
\end{figure}

Figure~\ref{fig:momentum} shows the momentum resolution for the tracks reconstructed by the \HS and \matching algorithms, in relative units, as function of the track momentum. It is below 1$\%$ for the majority of the tracks. If in addition a new dedicated \texttt{Parametrised Kalman Filter} algorithm is executed in the sequence\footnote{This algorithm performs a full track fit in all three tracking subdetectors, the VELO, UT and SciFi.}, the resolution improves to less than 0.5$\%$ across the whole momentum range. This algorithm builds on the parametrised Kalman filter of Ref.~\cite{Billoir_2021}, and its \hltone implementation will be described in a separate paper, currently in preparation.

\subsection{Tracking efficiency}
\begin{figure}[h!]
     \centering
     \begin{subfigure}{0.5\textwidth}
         \centering
         \includegraphics[width=\textwidth]{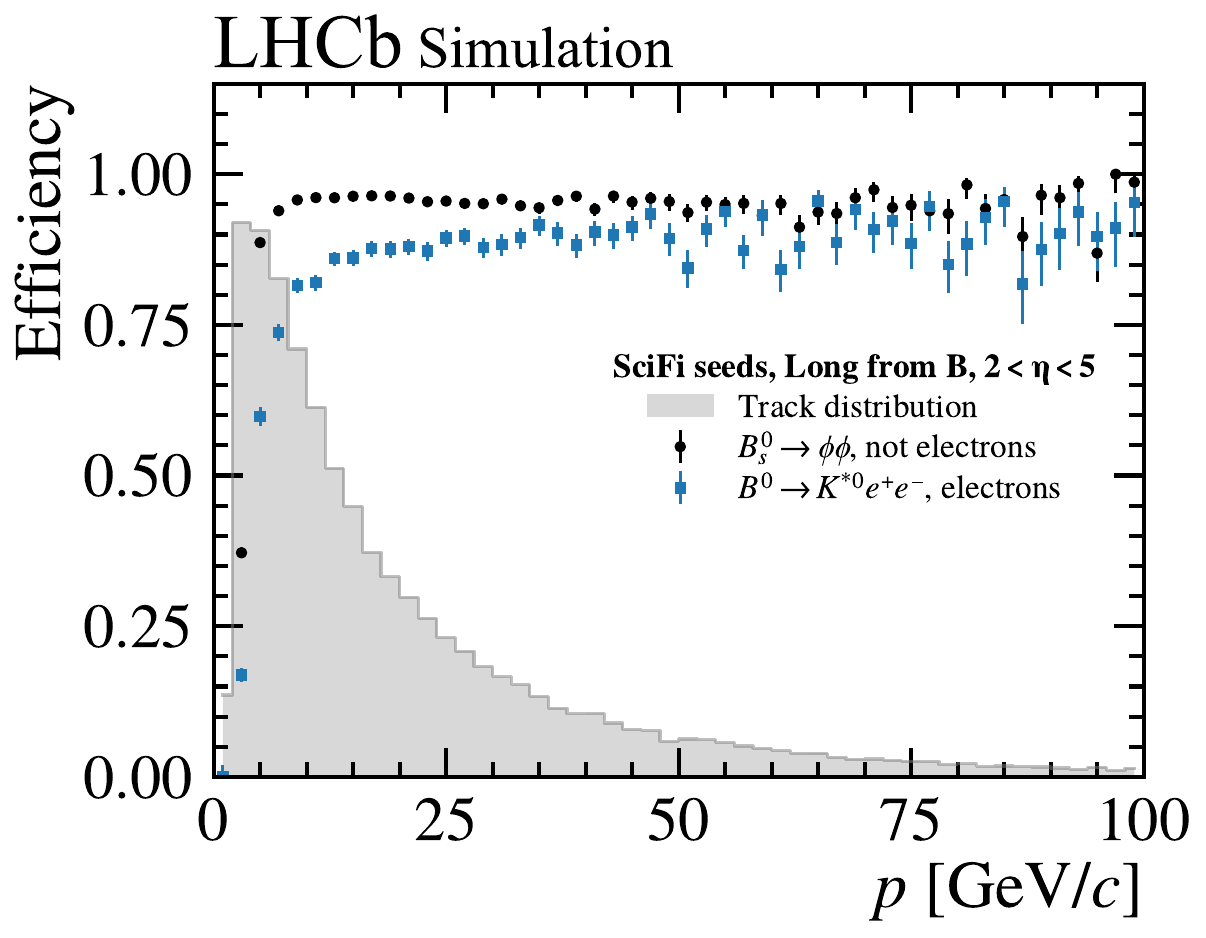}
         \caption{}
         \label{fig:eff_pt_hs}
     \end{subfigure}\hfill\begin{subfigure}{0.5\textwidth}
         \centering
         \includegraphics[width=\textwidth]{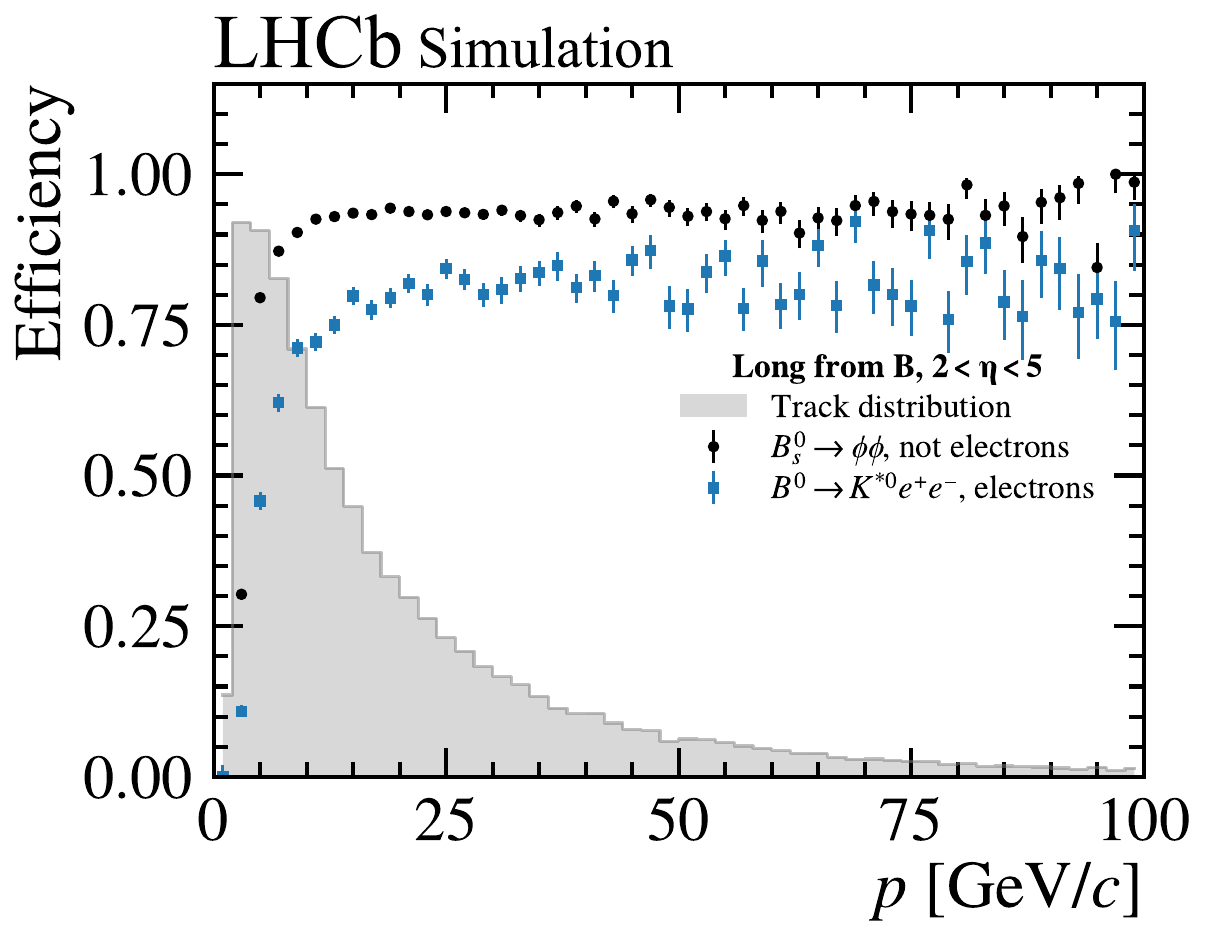}
         \caption{}
         \label{fig:eff_pt_tm}
     \end{subfigure}

        \caption{Track reconstruction efficiency as a function of momentum, $p$, for (top) seeds reconstructed by the \HS algorithm and (bottom) Long tracks reconstructed by the \texttt{Matching} algorithm. The track reconstruction efficiency in the pseudorapidity coverage of the LHCb detector, 2 $< \eta <$ 5, is shown with respect to reconstructible
        (black) non-electron and (blue) electron tracks
        passing through the VELO, UT and SciFi detectors and produced from $B$ decays. Reconstructibility criteria are defined according to Ref.~\cite{reco}.}
        \label{fig:eff_hs_tm }
\end{figure}

The efficiency of the \HS and \matching algorithms is measured using simulated $B_s^0 \to \phi \phi$ and $B^0 \to K^\ast e^+ e^-$  samples.  The $\phi$ particle from $B_s^0 \to \phi \phi$ decays into two kaon particles of opposite sign, which is representative for any reconstructed hadron. The efficiency as function of the transverse momentum, $p_T$, is shown in Fig.\ref{fig:eff_pt_hs} for \HS and \ref{fig:eff_pt_tm} for \matching algorithms, respectively. For high-$p_T$ hadrons the efficiency is larger than 90$\%$ (95$\%$) for \matching (\HS) algorithm in the LHCb pseudorapidity range ($2 < \eta < 5$). In the case of electrons the efficiency is reduced by about $10\%$ due bremsstrahlung losses before the magnet\footnote{
Electrons emit bremsstrahlung radiation as they traverse the detector material, which causes them to lose energy and appear to have lower momentum in the tracking system.}. The efficiencies as function of other variables such as the total momentum, track multiplicity and pseudorapidity distributions have also been studied, showing the expected behavior in the range of interest.  

\subsection{Ghost rate}
\begin{figure}[h!]
    \includegraphics[width=0.5\textwidth]{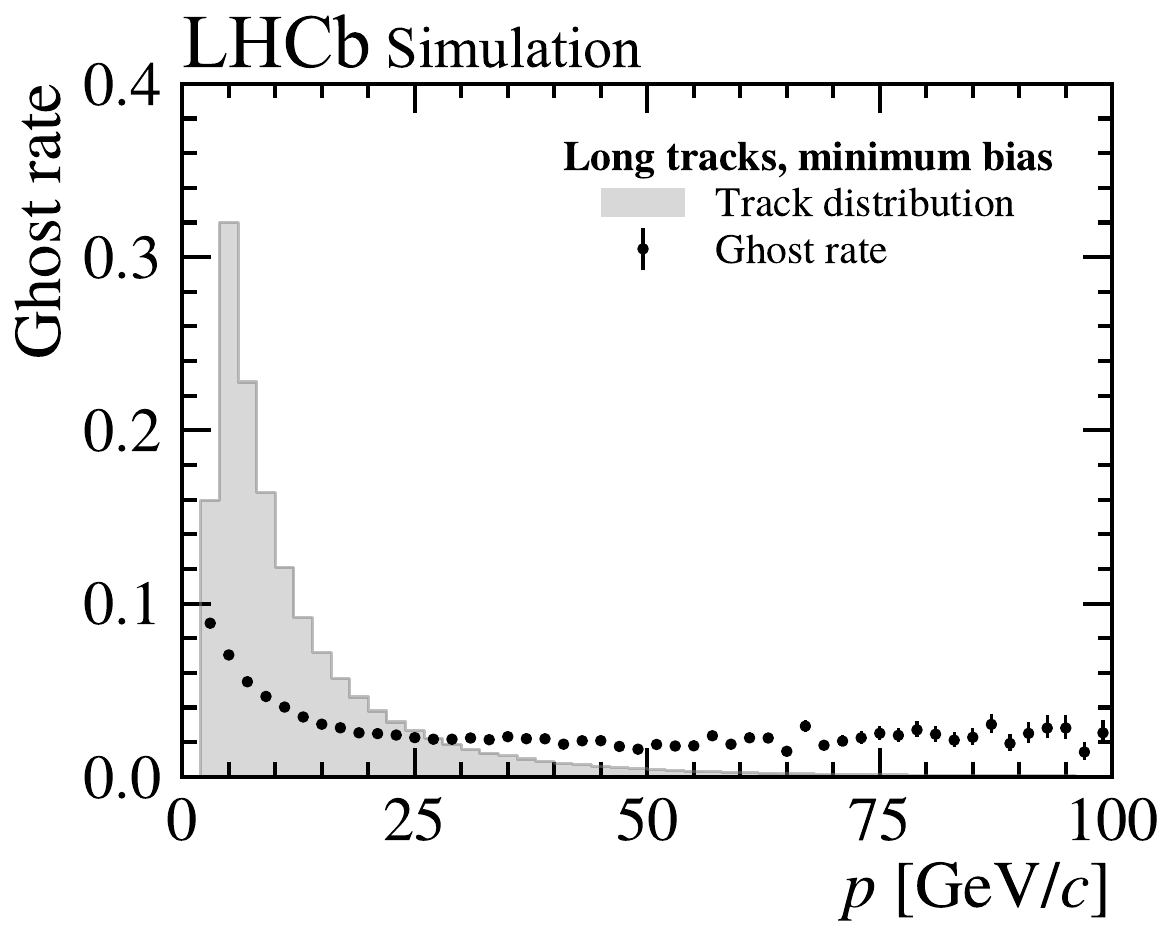}
    \caption{Ghost rate for HLT1 long tracks from $B$ decays as function of transverse momentum \pt. The plot shows reconstructed tracks from 10000 simulated $B_s^0 \to \phi \phi$ events where the \HS + \matching approach is used for the reconstruction.}
    \label{fig:ghost}
\end{figure}

Reconstructed tracks that cannot be attributed to any real particle are called \textit{ghost tracks} and are suppressed using a dedicated algorithm as explained in Sec.~\ref{sec:ghost-killing}.   
Figure~\ref{fig:ghost} shows the ghost rate for long tracks (hadrons) from $B$ decays as function of the transverse momentum. The threshold applied to the NN to reduce those fake tracks is 0.8, that is, removing all tracks with a 50$\%$ or more probability of being a ghost.   
With this condition the ghost rate is 8$\%$ for lowest $p_T$ tracks, but below 2$\%$ in most of the $p_T$ spectrum. 

\subsection{Energy consumption}
\begin{figure}
    \centering
\includegraphics[width=0.45\textwidth]{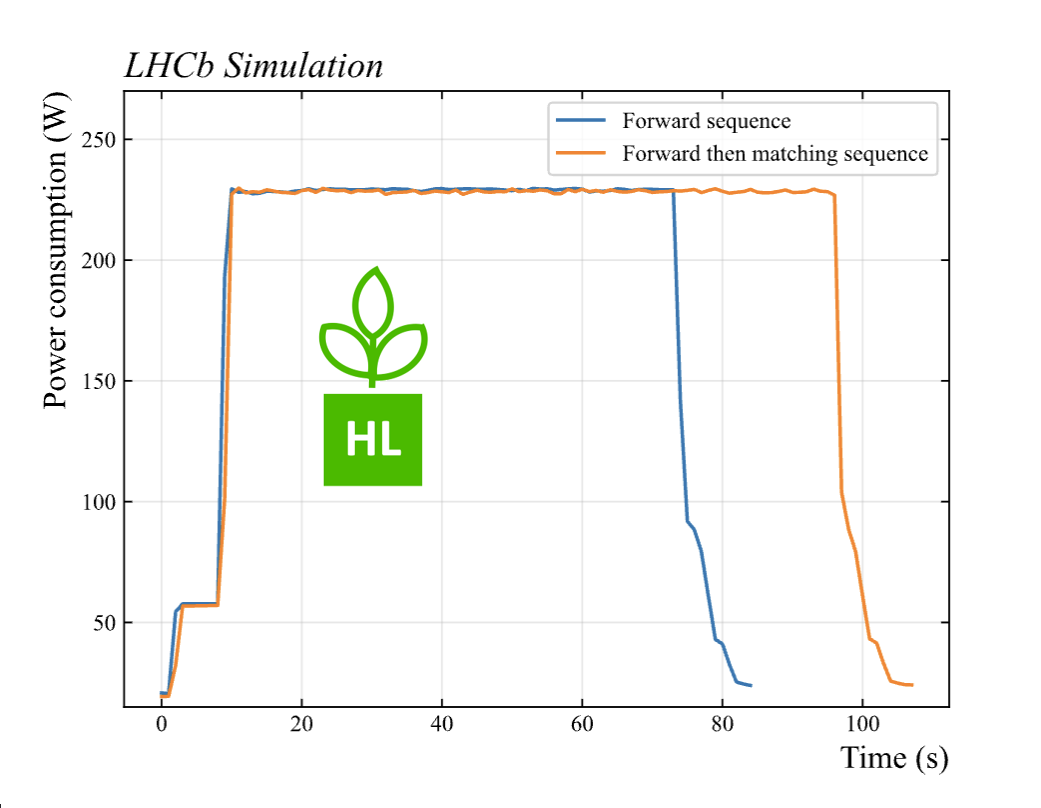}
    \caption{Energy consumption measured in a GPU RTX A5000 card for the ~\texttt{Forward} and \HS + \matching algorithms used in Allen to reconstruct long tracks. The increase in the energy consumption when adding \HS + \matching to the Allen sequence is 0.68\,mJ/event.}.    
\label{fig:power}
\end{figure}

Quantifying the energy consumption of the algorithms provides insights into the computational efficiency and sustainability, enabling optimization of resource usage. 
Figure~\ref{fig:power} shows the energy consumption vs. time of the \HS and \matching sequence in a RTX A5000 GPU card, as compared to the \texttt{Forward} algorithm. These measurements are performed by using a Nvidia DCGM driver~\cite{cuda-dcgm}. The effect of the inclusion of the \HS and \matching algorithms is a moderate increase of 0.68\,mJ/event. Both algorithms are properly optimised and make use of all the GPU resources, as indicated by the reach of the thermal design power (TDP) of the device.  

\section{Physics impact} 
\label{sec:physics}

\begin{figure}[h!]
    \centering
\includegraphics[width=0.48\textwidth]{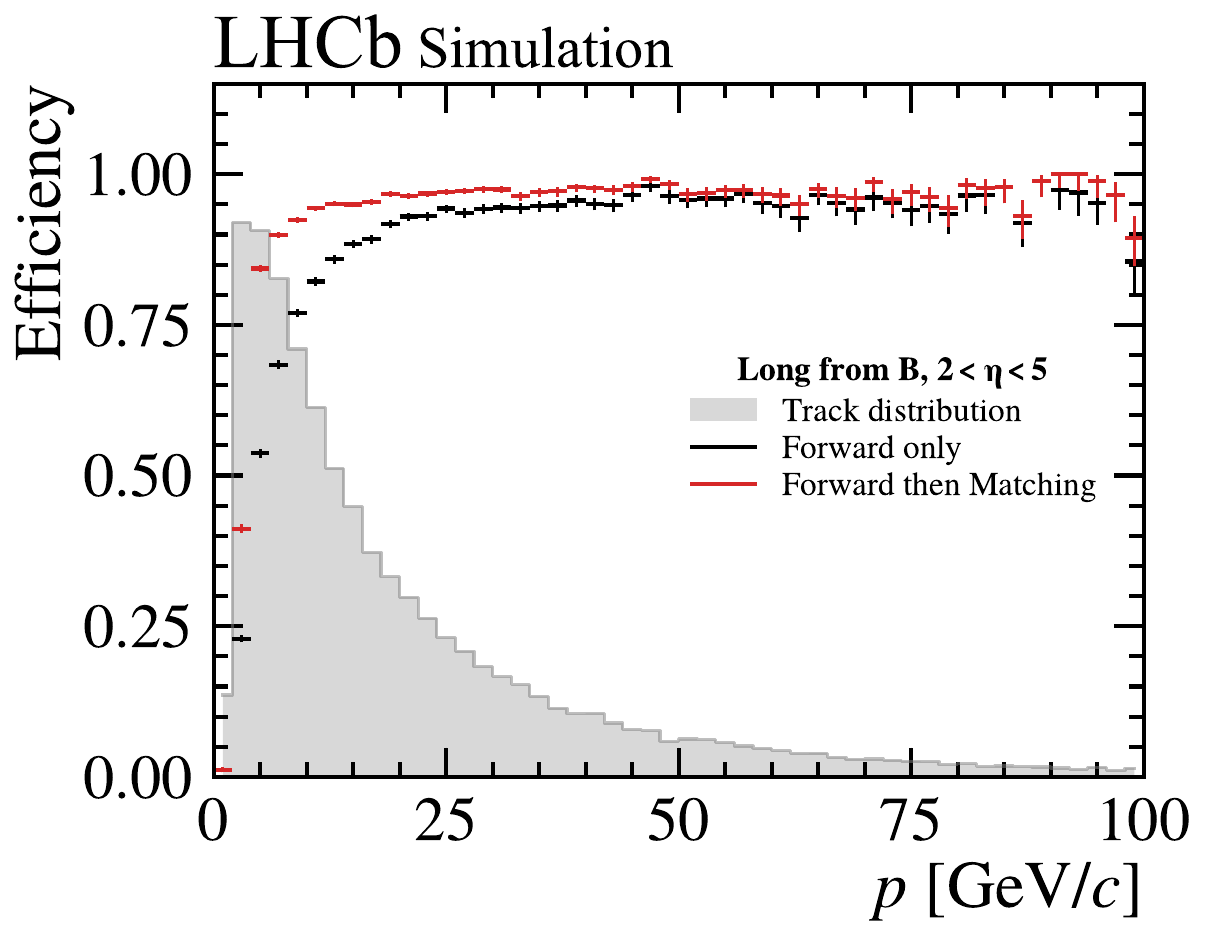}
    \caption{Track reconstruction efficiency as a function of momentum $p$ for long tracks from $B$ decays within $2 < \eta < 5$, reconstructed by the \texttt{Forward} algorithm alone (black) and by the \texttt{Forward} sequence followed by the \matching algorithm (red). The grey histogram shows the track momentum distribution. The \matching algorithm recovers low-momentum tracks not reconstructed by the \texttt{Forward} algorithm, increasing the physics potential of the first high-level trigger.}
\label{fig:comparison}
\end{figure}

The \HS and \matching algorithms are validated using the \texttt{Forward} algorithm as reference, to ensure that assumptions made during the reconstruction procedure such as the hypothesis for all seeds to originate from $x=y=0$, or the finite size of tolerance windows, do not bias the selection efficiency~\cite{forward}.  
Figure~\ref{fig:comparison} shows the effect of adding the \matching algorithm after the \texttt{Forward} execution, obtained from simulated $B_s^0\to\phi\phi$ events. The efficiency remains high for high-momentum tracks while improving at low momentum. Since the transverse momentum requirement imposed in the HLT1 \texttt{Forward} tracking algorithm is not present in the \matching, the new algorithms recover the reconstruction of low-momentum tracks that were previously excluded due to computational constraints. One should note that as compared with Ref.~\cite{forward}, here the running conditions of both algorithms have been tuned together to maximise the global efficiency. 

\begin{figure}[h!]
    \centering
\includegraphics[width=0.49\textwidth]{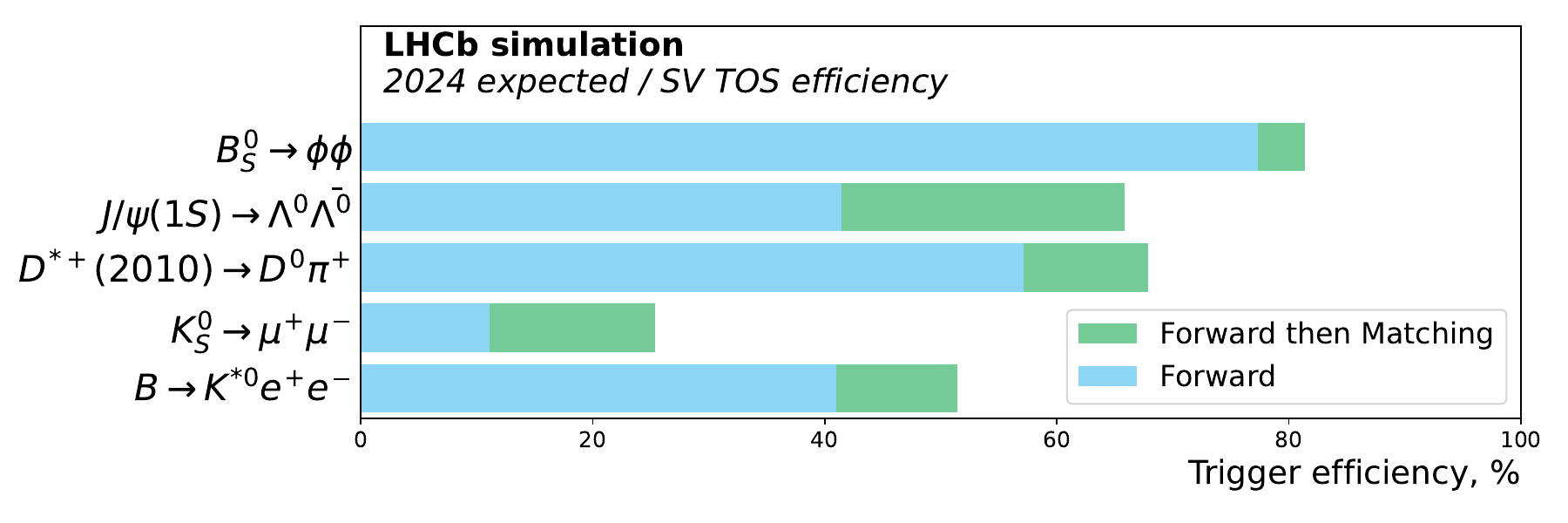}
    \caption{Increase of the physics efficiency when adding \HS and \matching to the Allen sequence for several decay channels using simulated events. The ${\rm D}^0$ decays into two \KS, with the \KS decaying to two opposite-charged pions. The $\Lambda$ decay to a proton-pion pair. The trigger requirement (TOS) on secondary vertices (SV) from signal events  is applied.}     
\label{fig:impact}
\end{figure}

This improvement significantly enhances the trigger’s efficiency and sensitivity to physics processes involving soft momentum particles, particularly those originating from charm and strange hadron decays, greatly improving the efficiency in corners of the phase space.
Figure~\ref{fig:impact} shows the increase in efficiency when adding the \HS and \matching algorithms to the \texttt{Forward} in the Allen sequence for several decay channels, assuming no bandwidth restrictions. No momentum criteria are applied for any of the algorithms. 
    
In addition, the \HS algorithm has been the starting point for the development of the \texttt{Downstream}~\cite{Kholoimov:2025cqe} algorithm of the Allen sequence, \ which has largely increased the efficiency for displaced tracks~\cite{Kholoimov:2025cqe, Gorkavenko_2024}.  

\section{Conclusion and prospects}\label{sec:conclusion}
The \HS and \matching algorithms, together with the \texttt{Downstream} algorithm in Ref.~\cite{Kholoimov:2025cqe} are the first
backward tracking strategy at the fully software-based first level trigger of LHCb, operating in real time on GPUs and processing 30 million collision events per second with a data throughput of 40\,Tb/s.
The procedure developed for the \matching algorithm largely increases the efficiency for low-momentum tracks that traverse the VELO detector when it is sequentially added to the previous reconstruction algorithm~\cite{forward}. In addition, the \HS algorithm enables the reconstruction of track candidates with no hits in the VELO detector~\cite{Kholoimov:2025cqe}.
This new backward strategy at HLT1 significantly impacts the physics program of LHCb during the Run~3. 

Additionally, the implementation of the HLT1 \HS has provided experience with SciFi standalone track reconstruction at the first trigger level in preparation for Run~4. The LHCb collaboration plans to install a dedicated FPGA-based device, the Downstream Tracker~~\cite{LHCb-tdr-025, Morello:2888549} which will perform the SciFi pattern recognition in hardware before the HLT, providing T tracks as an additional detector readout. 
The experience gained from operating the GPU-based \HS and \matching during Run~3 directly informs the design requirements and expected performance for Run~4 and 
demonstrates the viability of the backward-tracking approach.

\hspace{1cm}



\section*{Acknowledgements}
We thank LHCb's Real-Time Analysis project for its support, for many useful discussions, and for reviewing an early draft of this manuscript. We also thank the LHCb computing and simulation teams for producing the simulated LHCb samples used to benchmark the performance of the algorithm presented in this paper. The development and maintenance of LHCb's nightly testing and benchmarking infrastructure which our work relied on is a collaborative effort and we are grateful to all LHCb colleagues who contribute to it. VVG, CA, and LC were supported by the European Research Council under Grant Agreement number 724777 ``RECEPT'' while this research was carried out.
AO, JZ, VK and VS acknowledge the support from the Spanish Ministry of Science and Innovation, via the project TED2021-130852B-I00, and from the US NSF cooperative agreement OAC-1836650
(IRIS-HEP) and the Simons Foundation. This work was supported by the Swiss National Science Foundation (SNSF). Additional support to LH was provided by SNSF Grant No. 221385.



\bibliographystyle{splncs04}  
\setcounter{NAT@ctr}{0} 
\bibliography{sn-bibliography2} 
\end{document}